\documentclass[useAMS,usenatbib]{mn2e}
\usepackage{graphicx}
\usepackage{color}
\usepackage{amsmath,amssymb}             
\def\jose{J. Fiestas}
\defcitealias{einsel99}{Paper~I}
\defcitealias{kim02}{Paper~II}
\defcitealias{kim04}{Paper~III}

\def\ARAA{Ann. Rev. Astron. Astroph.}

\def\ApJ{Astrophys. J.}
\def\ApJL{Astrophys. J. Lett.}

\def\AJ{Astron. J.}

\def\PASJ{Publ. Astron. Soc. Japan}

\def\MN{Mon. Not. Royal Astron. Soc.}

\title[Dynamical evolution of rotating dense stellar systems with embedded black holes]{Dynamical evolution of rotating dense stellar systems with embedded black holes}
\author[\jose, R. Spurzem]{
\jose$^{1}$\thanks{E-mail:fiestas@ari.uni-heidelberg.de} and
 ~R.Spurzem$^{1,2,3}$\thanks{E-mail:spurzem@ari.uni-heidelberg.de}\\
$^{1}$National Astronomical Obervatories of China, Chaoyang District, Beijing 100012, China\\
$^{2}$The Kavli Institute for Astronomy and Astrophysics at Peking University\\
$^{3}$Astronomisches Rechen-Institut, Zentrum f\"{u}r Astronomie der Universit\"{a}t Heidelberg, M\"{o}nchhofstraße 12-14,   D-69120 Heidelberg, Germany
}
\begin{document}
\pagerange{\pageref{firstpage}--\pageref{lastpage}} \pubyear{2010}
\maketitle
\label{firstpage}
\begin{abstract}
Evolution of self-gravitating rotating dense stellar systems (e.g. globular clusters, galactic nuclei) with embedded black holes is investigated. The interaction between the black hole and stellar component in differential rotating flattened systems is followed. The interplay between velocity diffusion due to relaxation and black hole star accretion is investigated together with cluster rotation using 2D+1 Fokker-Planck numerical methods. The models can reproduce the Bahcall-Wolf solution $f \propto E^{1/4}$ ($n \propto r^{-7/4}$) inside the zone of influence of the black hole. Gravo-gyro and gravothermal instabilities conduce the system to a faster evolution leading to shorter collapse times with respect to the non-rotating systems. Angular momentum transport and star accretion support the development of central rotation in relaxation time scales. We explore system dissolution due to mass-loss in the presence of an external tidal field (e.g. globular clusters in galaxies).
\end{abstract}
\section{Introduction}
Observational analysis of globular clusters (GCs) has been considerably improved in the last years thanks the Hubble Space Telescope (HST), (cf. e.g. \citealt{piotto02,rich05,beccari06,georg09}). They have been used to obtain luminosity functions and derived mass functions, color-magnitude diagrams (CMDs) and population and kinematical analysis leading to a better understanding of their evolutionary processes. At the same time, new questions have been opened due to the, in the past unimpressive, complexity of this stellar systems. Nevertheless, analysis of rotation in these systems has been rarely carried on. On the other side, there are observational evidences for the existence of intermediate-mass black holes (IMBHs) in GCs, although their origin is as yet not very clear. Local core collapsed GCs are expected to harbor IMBHs due to their high central densities, but regarding its detection, it was argued that 'non-collapsed' projected density profiles which evolve harboring IMBHs fit well by medium-concentration King models \citep{baum05}. \cite{ger02,ger03} reported the kinematical study (based on HST spectra) of the central part of the collapsed GC M15. They proposed the presence of an IMBH ($M_{\rm bh}=3.9 \cdot 10^3 M_\odot$) in the central region of M15. A single or binary IMBH could account for the net rotation observed in the center of M15 \citep{geb00,ger02,miller04,kiselev08}. \cite{maccarone08} shows the possible presence of an IMBH in NGC 2808 with a mass of $\sim 2.7 \times 10^3 M_\odot$ and \cite{noyola08} have reported about the BH in omegaCen with a mass of $\sim 10^4 M_\odot$ . However, \cite{baum03} have shown through self-consistent $N$-body computations treating stellar evolution and with a realistic IMF, that the core collapse profile of a star cluster with an unseen concentration of neutron stars and heavy-mass white dwarfs can explain the observed central rise of the mass-to-light ratio (see also \citealt{mcnam03}). Similarly, a dense concentration of compact remnants might also be responsible for the high mass-to-light ratio of the central region of NGC 6752 seen in pulsar timings \citep{ferr03,colpi03}. Outside our own galaxy, \cite{geb02a,geb05,zaharija08} have reported evidence for a $20000\,M_\odot$ BH in the M31 globular cluster G1.

Chandra and XMM-Newton observations of ultraluminous X-ray sources (ULXR) give also evidence for the existence of IMBHs in dense star clusters, which are often associated with young star clusters and whose high X-ray luminosities in many cases suggest a compact object mass of at least $10^2 M_\odot$ \citep{ebi01,miller03}. Furthermore, some of the ULXR sources detected in other galaxies are may be accreting IMBHs (e.g., \citealt{miller04}), although the majority could be likely stellar-mass black holes \citep{king01,rappap05}.

On the other side, the centers of most galaxies embed massive BHs. It has been evidenced by HST measurements in the last years, and since theoretical modeling of measured motions request for the presence of central compact dark objects with a mass of $\sim 10^6 \ {\rm to} \ 10^9 M_\odot$ \citep{ferra01,geb02b,pink03,kor04}. Ground-based IR observations of the fast orbital motions of a few stars in the Milky Way have lead to the detection of a $3 - 4 \times 10^6 M_\odot$ BH in its center \citep{schoed03,ghez04,eckart04}. Moreover, BH demographics have lead to correlations between the BH mass and the luminosity of its host bulge or elliptical galaxy \citep{kor95}, and between BH mass and the velocity dispersion of its host bulge, as $M_{\rm bh} \propto \sigma^\alpha$ \citep{ferra00}. This leads to a strong link between BH formation and the properties of the stellar bulge, like the formation of density cusps (Bahcall-Wolf solution), which have been investigated by \cite{schoedel08}

Dynamical modeling of collisional stellar systems (like galactic nuclei, rich open clusters, and rich galaxy clusters) still possess a considerable challenge for both theory and computational requirements (in hardware and software). On the theoretical side the validity of certain assumptions used in statistical modeling based on the Fokker-Planck (FP) and other approximations has not been fully investigated. Stochastic noise in a discrete $N$-body system and the impossibility to directly model realistic particle numbers with the presently available hardware, are a considerable challenge for the computational side (but see \cite{berczik06})

While all work known to the authors at this moment concentrates on self-gravitating star clusters, the improvement of our knowledge and methods in the field of rotating dense stellar systems is extremely important for galactic nuclei, too, where a central star-accreting black hole comes into the game (some stationary modeling exists, such as \citealt{duncan83}, \citealt{quinlan90}, \citealt{murphy91}, \citealt{freitag02}). Direct integration of orbits ($N$-Body method) has been applied to the problem \citep{gultekin04,baum05}. However, $N$-Body simulations only provide a very limited number of case studies, due to the enormous computing time needed even on the GRAPE computers. Moreover, recent investigations show that in young dense clusters, supermassive stars may form through runaway merging of main-sequence stars via direct physical collisions, which may then collapse to form an IMBH. The collision rate will be greatly enhanced if massive stars have time to reach the core before exploding as supernovae \citep{port04,gurk06,frei06}. Therefore it is very urgent to develop reliable approximate models of rotating star clusters with black hole, which is subject of the present work.

A 2D FP model has been worked out for the case of axisymmetric rotating star clusters \citepalias{einsel99,kim02}. Here, the distribution function is assumed to be a function of energy $E$ and the $z$-component of angular momentum ($J_z$) only; a possible dependence of the distribution function on a third integral is neglected. As in the spherically symmetric case the neglect of an integral of motion is equivalent to the assumption of isotropy, here between the velocity dispersions in the meridional plane ($\varpi$ and $z$ directions); anisotropy between velocity dispersion in the meridional plane and that in the equatorial plane ($\varphi$-direction), however, is included.

We realize that the evolutionary models provided by us for rotating dense stellar systems are difficult to use for direct comparisons with observations, because they are not easily analytically describable. But they are the only ones which fully cope with {\it all} observational data available nowadays (full 3D velocity data, including velocity dispersions in $\varpi$ and $\varphi$-direction, rotational velocity, density, all as full 2D functions of $\varpi$ and $z$, see Fiestas et al. 2006). No other evolutionary model exists so far which is able to provide this information. With the advent of our new post-collapse and multi-mass models \citepalias{kim02,kim04} and the inclusion of stellar evolution and binaries (work in progress) we will be able to deliver even more interesting results. Already the existing $N$-body study \citep{ardi05} shows that rotation not only accelerates the collisional evolution (but see Ernst et al. 2007) but also leads to an increasing binary activity in the system.

A description of the method used in the present work is made in Section \ref{sec:2}, Section \ref{sec:3} describes the initial configurations of the models and numerical tests of the code, Section \ref{sec:4} presents the main results in the isolated case first reproducing the spherically symmetric model (no rotation), and secondly, giving a description of rotational behavior of axisymmetric systems in relaxation time scales, emphasizing the interplay between the dynamical evolutionary processes. Section \ref{sec:5} explores the system dissolution due to the tidal field of a parent galaxy. Section \ref{sec:6} gives the conclusions and further plans.
\section{Theoretical model}
\label{sec:2}
\subsection{Equations and assumptions}
\label{sec:21}

The pioneering work of \cite{goodman83}, in his unpublished thesis, and the further development of the Fokker-Planck method made by \cite{einsel99} and \cite{kim02,kim04}, have brought the treatment of the axisymmetric rotating case to a newly state of interest, which follows the evolution of self-gravitating rotating systems driven by relaxation effects and its consequences for the stellar redistribution and shape of the system.

Evolution of the distribution function $f (\vec{r},\vec{v})$ \footnote{a proper definition of $f$ corresponding to the initial conditions for this study is given in Eq.~\ref{king}} of stars in phase space $(\vec{r},\vec{v})$ under the influence of the potential $\phi(\vec{r}) $ is described by the Boltzmann-equation
\begin{eqnarray}
\frac{\partial f}{\partial t} + \vec{v} \cdot \vec{\nabla}_{\rm r} + \frac{\vec{F}}{m} \cdot  \vec{\nabla}_{\rm v} f = (\frac{\partial f}{\partial t})_{\rm coll}
\label{boltzeq}
\end{eqnarray}
with space and velocity coordinates, $\vec{r}$ and $\vec{v}$ respectively. The force $\vec{F}=-m \vec{\nabla}_{\rm r} \phi$ is applied on stars of mass $m$. The term on the right side of Eq.~\ref{boltzeq} takes into account the changes in $f$ due to collisions (not real collisions but stellar scatterings which cause deviations in the orbits). The collision term is given through the -local- Fokker-Planck approximation
\begin{eqnarray} 
\Bigl(\frac{\partial f}{\partial t}\Bigr)_{\rm coll} =  - \frac{\partial}{\partial \upsilon^\mu}   (f \langle \Delta \upsilon^\mu \rangle) + \frac{1}{2} \frac{\partial^2}{\partial\upsilon^\mu  \partial\upsilon^\nu} (f\langle \Delta \upsilon^\mu\Delta \upsilon^\nu\rangle)
\label{collterm}
\end{eqnarray}
where $\mu$ = 1,2,3 and $\nu$ = 1,2,3 (tensor notation). $\upsilon^\mu$ gives the velocity in Cartesian coordinates.  The first order diffusion coefficients $\langle \Delta \upsilon^\mu \rangle$ describe the dynamical friction, while the second order ones $\langle \Delta \upsilon^\mu \Delta \upsilon^\nu\rangle$ give the real velocity diffusion.

In order to obtain the solution of the Fokker-Planck equation following assumptions are taken into account:
\begin{itemize}
\item cluster evolution time scales are in following relation:
\begin{eqnarray}
t_{\rm dyn} \ll t_{\rm rh} \ll t_{\rm cl}
\end{eqnarray}
where $t_{\rm cl}$ represents the cluster age. $t_{\rm rh}$ is the half-mass relaxation time, following \cite{spitzer71}:
\begin{eqnarray}
t_{\rm rh} = \frac{0.138 \sqrt{N r_{\rm h}^3}}{\sqrt{Gm} \ ln\Lambda}
\label{trh}
\end{eqnarray}
$N$ is the number of particles (stars), $G$ the gravitational constant, $m$ the mean stellar mass, $ln\Lambda$ is the Coulomb logarithm ($\Lambda=0.4N$ is used here) and $r_{\rm h}$ the half-mass radius. The dynamical time is given by
\begin{eqnarray}
t_{\rm dyn} = 1.58 \sqrt{\frac{r_{\rm h}^3}{GM}}
\label{tdyn}
\end{eqnarray}
where $M$ is the total mass of the cluster.\\
The system evolves slowly through diffusion in a sequence of virtual equilibrium states. In a time $t_{\rm rh}$ information about the initial configuration is lost due to relaxation. A prove of this statement is shown in Section \ref{sec:32}.

\item the solution is given for small-angle scatterings ($\Delta \upsilon/ \upsilon \ll 1$), i.e., for changes of $\vec{\upsilon}$ to $\vec{\upsilon} + \Delta\vec{\upsilon}$.
\item there is no correlation between collisions (like in three-body collisions), which could be important for the energy generation in the core that can reverse the collapse.
\item no binaries and stellar evolution are considered. Thus, binary heating due to 3-body encounters is neglected. Note that binary heating can reverse collapse \citep{hut85,mcmillan90,kim02}
\item We neglect any recoil of the BH as a result of accretion and three-body interactions. If these were to be taken into account the initial BH seed would end up kicked out of the system well before it has a chance to significantly grow in mass. Realistic effects on the dynamical evolution are being studied by using NBody realisations of the systems, to be published in a forthcoming paper.
\item the initial BH mass ($M_{\rm bh_i}$) is much smaller than the cluster mass $M_{\rm cl}$
\item The distribution of stars is represented by an equal-mass particle system, which is initially axisymmetric in space and is able to develop anisotropy in velocity space. No stellar spectrum is included in this model, with the aim to test the model without large complexity.
\end{itemize}

The classical isolating integrals of a general axisymmetric potential $\phi$, in cylindrical coordinates ($\varpi,z$), are the energy per unit mass:
\begin{eqnarray}
E=\frac{1}{2}v^2 + \phi_{\rm cl}(\varpi,z) + \phi_{\rm bh}(\varpi,z)
\end{eqnarray}
\noindent
where $\phi_{\rm cl}(\varpi,z)$ is the potential of the stellar system and $\phi_{\rm bh}(\varpi,z) = - GM_{\rm bh}/r$ is the BH-potential ($r^2=\varpi^2+z^2$); and the component of angular momentum along the z-axis per unit mass, given by
\begin{eqnarray}
J_z=\varpi \vec{v} \hat{e}_\varphi
\end{eqnarray}
\noindent
$v_\varphi=\vec{v} \hat{e}_\varphi$ is the velocity component in azimuthal direction.
$E$ and $\phi$ are negative for all particles.

Conservation of $E$ and $J_{\rm z}$ is used in the solution of the Fokker-Planck equation, which becomes a non-linear second order integro-differential equation (the diffusion coefficients of Eq.~\ref{collterm} are expressed in terms of integrals over the local field star velocity distribution function). This integrals are given by the Rosenbluth potentials (Rosenbluth et al., 1987). A derivation of the diffusion coefficients in terms of $E$ and $J_{\rm z}$ can be found in \cite{einsel99}.

In axisymmetric systems, although $f$ can be approximately representable as a function of $E$, $J_z$ and $t$ (except for very special forms of the potential), numerical evidence demonstrates that axisymmetric potentials can support orbits which have three integrals of motion: $E$, $J_{\rm z}$ and a third integral commonly designated $I_3$. That is, the typical orbit does not spread uniformly over the hypersurface in phase space defined by its $E$ and $J_z$ but is confined to lower-dimensional subset ('non-ergodic' orbits on their $EJ_z$ surfaces). A solution of the orbit-averaged FP equation in energy-momentum space may represent an artificial case of a true point-mass system, since in the axisymmetric potential a third integral of motion could restrict particle motion in phase space \citep{goodman83}. Since on one side, the inner parts of the cluster are dominated by relaxation effects and the third integral can be neglected, due to the efficiency of diffusion in these regions; on the other side, the outer region of the cluster can be strong influenced by the third integral, as radially biased anisotropy dominates this region.

In the present study, non-ergodicity on the hypersurface (given by $E$ and $J_{\rm z}$) is neglected due to any third integral $I_3$. The potential close to the BH is spherically symmetric ($\sim 1/r$), and $I_3$ could be fairly approximated by $J^2$, since less radial orbits are expected in this regions \citep{amaro04,baum04}, which are preferentially disrupted by the BH. The angular momentum  $J_{\rm z}$ is here good represented by its maximum value ($ J_{\rm z}^{\rm max}$). Nevertheless, possible existing meridional circular orbits can not be distinguished by our model and would be treated as radial orbits (for example for their accretion).

In terms of integrals of motion, the Boltzmann equation (Eq.~\ref{boltzeq}) is expressed in the axisymmetric system as:
\begin{eqnarray}
\frac{\partial f}{\partial t} + \frac{\partial\phi}{\partial t}\frac{\partial f}{\partial E} = \Bigl(\frac{\partial f}{\partial t}\Bigr)_{\rm coll}
\label{fpej}
\end{eqnarray}
the dependence on $J_{\rm z}$ is given implicitly by $\phi$; and the collisional term of Eq.~\ref{collterm} can be expressed in terms of $E$ and $J_{\rm z}$ as
\begin{eqnarray*}
 \Bigl(\frac{\partial f}{\partial t}\Bigr)_{\rm coll} = \frac{1}{V} \Bigl\lbrack - \frac{\partial}{\partial E}(\langle \Delta E\rangle V f)- \frac{\partial}{\partial J_{\rm z}}(\langle \Delta J_{\rm z}\rangle V f) +
\end{eqnarray*}
\begin{eqnarray*}
+ \frac{1}{2} \frac{\partial^2}{\partial E^2} (\langle (\Delta E)^2\rangle V f)+ \frac{1}{2} \frac{\partial^2}{\partial J_{\rm z}} (\langle\Delta E \Delta J_{\rm z}\rangle V f)+
\end{eqnarray*}
\begin{eqnarray}
 \frac{1}{2} \frac{\partial^2}{\partial J_{\rm z}^2} (\langle (\Delta J_{\rm z})^2\rangle V f)  \Bigr\rbrack
\label{fpapprox}
\end{eqnarray}
with the volume element in velocity space given by $V = \frac{2\pi}{\varpi}$.

The vast dynamical parameter range of relaxed and unrelaxed cluster systems (as presented in \cite{fiestas06}) was specially treated applying appropriate boundary conditions at the inner potential cusp of the BH and the outer cluster tidal boundary (in the presence of a parent galaxy). A double-logarithmic $(\varpi,z)$ space grid is used, and the Fokker-Planck equation is written in a dimensionless flux form by introducing the dimensionless energy
\begin{eqnarray}
X(E) \equiv \ln(\frac{E}{2\phi_c-E_0-E});
\label{xgrid}
\end{eqnarray}
$E_0$ is a characteristic energy, which allows a higher resolution at higher energy (and low angular momentum) levels, as well as in the outer parts of the system (halo), where the proportionality $X-ln|E|$ improves the spacing of the radii of circular orbits with given energies in the direction of the tidal boundary. The dimensionless angular momentum is given by
\begin{eqnarray}
Y(J_z,E) \equiv \frac{J_z}{J_z^{max}}
\label{ygrid}
\end{eqnarray}

Each time step $r_{\rm circ}(E)$ and $J_{\rm z}^{\rm max}(E)$ are determined in the equatorial plane from the evolving potential by a simple Newton--Raphson scheme, using the relations:
\begin{eqnarray}
(E-\phi(\varpi_{\rm circ},z=0))= \frac{1}{2} \varpi_{\rm circ} \frac{\partial \phi}{\partial \varpi}
\end{eqnarray}
in order to get $\varpi_{\rm circ}$ (or $r_{\rm cric}$, due to $z=0$), and computing $J_{\rm z}^{\rm max}(E)$, using:
\begin{eqnarray}
(J_{\rm z}^{\rm max}(E))^2 = r_{\rm circ}^3 \frac{\partial \phi}{\partial r}
\end{eqnarray}

\subsection{Diffusion and loss-cone accretion}
\label{sec:22}

Given the values of $E$ and $J_{\rm z}$, the orbit average of the Fokker-Planck equation in the form of Eq.~\ref{fpej} is obtained by integrating it over an area $P(E,J_{\rm z},t)$ of the hypersurface in phase space, given by
\begin{eqnarray}
P(E,J_{\rm z},t)=4 \pi^2 \int \int_{A(E,J_{\rm z})} d\varpi dz
\label{p}
\end{eqnarray}
This weighting factor gives also the number of stars in the system taking part in the diffusion, as
\begin{eqnarray}
N(E,J_{\rm z},t)= P(E,J_{\rm z},t) \cdot f(E,J_{\rm z},t)
\end{eqnarray}
$A(E,J_{\rm z})$ is given by the intersection of the hypersurface with the $\varpi z$-plane, where the sum of the squares of the velocity components are non-negative:
\begin{eqnarray}
A(E,J_{\rm z}) \equiv \Bigl\lbrace (\varpi z) \ \ \vert \ \ \frac{1}{2} \vec{v}_\varpi^2 + \vec{v}_z^2 = E-\phi- \frac{J_{\rm z}^2}{2\varpi^2} \geq 0 \Bigr\rbrace
\label{area}
\end{eqnarray}
The condition (\ref{area}) is rastered numerically in the code by given $E$ and $J_{\rm z}$.

In a general axisymmetric potential, almost none of the orbits are closed, so that the orbital period is not well defined. There exist two different epicycle periods, one each for oscillations in the $\varpi$ and $z$ directions. The orbit average is taken over a time that is larger than both and is the time required for the orbit to spread uniformly over the area $A(E,J_{\rm z})$, only because of encounters, i.e. in a relaxation time scale (if the third integral is well conserved).

The Fokker-Planck equation is solved numerically in flux conservation form, following :
\begin{eqnarray}
\frac{df}{dt} = \frac{1}{p} (-\frac{\partial F_X}{\partial X}-\frac{\partial F_{\rm Y}}{\partial Y})
\label{fpflux}
\end{eqnarray}
$p$ is the phase volume per unit $X$ and $Y$, with particle flux components in the $X$ and $Y$ directions:
\begin{eqnarray}
F_X=-D_{\rm XX} \frac{\partial f}{\partial X} -D_{\rm XY} \frac{\partial f}{\partial Y} -D_X f;
\label{fluxe}\\
F_{\rm Y}=-D_{\rm YY} \frac{\partial f}{\partial Y} -D_{\rm YX} \frac{\partial f}{\partial X} -D_{\rm Y} f
\label{fluxj}
\end{eqnarray}
The orbit-averaged flux coefficients $D_{\rm ii}$ are derived from the local diffusion coefficients and transformed to dimensionless variables $D_{\rm X},D_{\rm Y},D_{\rm XX},D_{\rm YY},D_{\rm XY}$ \citep{einsel99}.

The loss-cone limit is defined by the minimum angular momentum for an orbit of energy E:
\begin{eqnarray}
J_{\rm z}^{min}(E) = r_d \sqrt{2(E-GM_{\rm bh}/r_d)}
\label{jzmin}
\end{eqnarray}
where $r_d$ is the disruption radius of the BH, calculated following \cite{fr76}:
\begin{eqnarray}
r_{\rm d} \propto r_{\ast}(M_{\rm bh}/m_{\ast})^{1/3}
\label{rdis}
\end{eqnarray}
$r_{\ast}$ and $m_{\ast}$ are the stellar radius and mass, respectively. Their adopted values are given in Section \ref{sec:31}.

The central potential cusp of an embedded massive BH disturbs the redistribution of stars due to collisional interactions. Thus, following assumptions are taken in order to develop the structural parameters of the cluster:

\begin{itemize}
\item a seed initial BH mass, which is much larger than a stellar mass, is calculated numerically using a first perturbation of the potential in the initial models (see also Section \ref{sec:31})
\item accretion is driven by angular momentum diffusion. A star is completely accreted if its $z$-component of angular momentum is less than $J_{\rm z}^{\rm min}$, which defines the loss-cone boundary. Energy diffusion for accretion is neglected because the changes in $E$ due to collisions are considered small in comparison to the changes in angular momentum \citep{cohn78}.

\item the distribution function vanishes for $J_{\rm z} > J_{\rm z}^{\rm max}$ and $J_{\rm z} < -J_{\rm z}^{\rm max}$.
\item the central BH grows slowly through accretion of stars leading to a new distribution $f(E,J_{\rm z})$ and a new $\phi(\varpi,z)$.

\item The unit of time is proportional to the relaxation time at the influence radius of the BH $r_{\rm a}$, defined as the radius, where the mass of the cluster equals $M_{\rm bh}$. The time step is given by $\Delta t = \xi(t) \tau_{\rm ra}$, where:
\begin{eqnarray}
\tau_{\rm ra} = \frac{0.338 \sigma_{\rm a}^3}{n_{\rm a} (Gm_\star)^2 ln\Lambda}  
\label{trx}
\end{eqnarray}
$\sigma_{\rm a}$ and $n_{\rm a}$ are the velocity dispersion and density evaluated at $r_{\rm a}$. $\xi(0)$ depends on the initial model and is increased from time to time  by a factor of 4/3 in order to have a fractional increasing of central density of between 2 and 4~\% per time step. In analogy with the computations of \cite{cohn79}, one Vlasov step, i.e. one recomputation of the potential, follows every Fokker-Planck (diffusion) step.
\end{itemize}

\begin{figure}
\includegraphics[width=\linewidth]{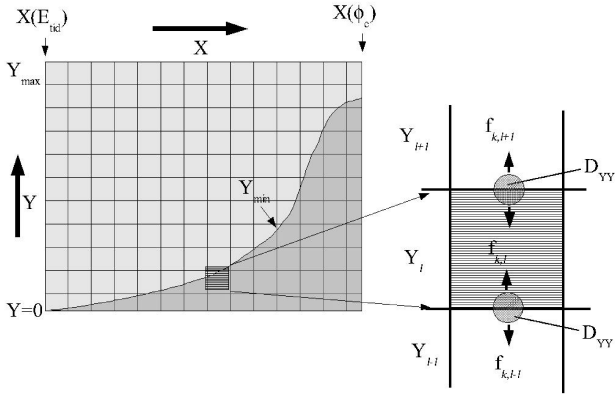}
\caption{\small Schematic diagram of the numerical $XY$-grid and definition of the loss-cone. Half of the grid is shown in the left side of the figure. The other lower half corresponds to negative values of Y and is symmetric with respect to the $Y=0$-axis. The dark shaded area represents the loss-cone in $X-Y$-space, and is limited by $Y_{\rm min}$. Stars are able to go into and out of it through angular momentum diffusion, as shown in the right side of the figure, for one grid cell.}
\label{losscone}
\end{figure}

A schematic diagram of the numerical $XY$-grid, as used for the solution method of the discretized Fokker-Planck equation, is shown in Fig.~\ref{losscone}. Energy limits are the central potential ($X(\phi_{\rm c})$) and the tidal energy ($X(E_{\rm tid})$), angular momentum limits are the maximum values of $Y$ in both directions ($Y= \pm 1$). For the purpose of this illustration, only the upper half of the grid is shown ($0 \leq Y<+1$), since the lower half ($-1<Y<0$) is symmetric with respect to the $Y=0$-axis. Angular momentum diffusion of stars into and out of neighboring cells is illustrated in the right part of the figure. The distribution function is defined in the center of each cell, while the diffusion terms are computed at the cell boundaries. As shown in Fig.~\ref{losscone} the fluxes at the Y=±1 boundaries are set to zero. Open boundaries are the loss-cone region and $E=E_{\rm tid}$ in the tidally limited models. The loss-cone is limited by $± Y_{\rm min}(E)$ and the limit of maximum energy ($E(\phi_c)$). The first derivative of f with respect to E is non-zero at the boundary of each grid cell and is evaluated just inside the boundary in order to obtain accurate escape fluxes. During the solution method, the whole grid is rastered and the angular momentum fluxes are saved each time step. In case of isolated models $X(E_{\rm tid})$ is set to 1/10 of the potential at the tidal radius of the corresponding King model. Evaporation of stars in isolated systems is neglected. In tidally limited systems, stars are able to escape the system through the tidal boundary at $X(E_{\rm tid})$, influenced by the potential field of the parent galaxy.

The flux term $F_{\rm Y}$, per unit energy and unit time across the angular momentum boundary, used to compute contribution to stellar accretion, is given by the second order angular momentum diffusion term in Eq.~\ref{fluxj}
\begin{eqnarray}
F_{\rm Y}=-D_{\rm YY} \frac{\partial f}{\partial Y}
\label{fluxj1}
\end{eqnarray}

The dimensionless change in $f(X,Y)$, due to diffusion in the inner/outer direction is then given in a discretized form by:
\begin{eqnarray}
\frac{\Delta f}{f} = \frac{\Delta t}{P} \frac{\Delta F_{\rm Y}}{\Delta Y} \frac{1}{f}
 \label{yfluxdisc}
\end{eqnarray}
where $\Delta t$ is the time step, $P(X,Y)$ the phase space volume and $\Delta F_{\rm Y}$ is the neto angular momentum flux in each cell. Energy fluxes are neglected for accretion since they are small in comparison to $F_Y$ \citep{cohn78}.

After redistribution of orbits, due to small-angle collisions, those with $Y \leq Y_{\rm min}$ lie in the loss-cone. Using the time scales of replenishment $t_{\rm in} \propto (Y^{\rm min})^2$ \footnote{the square root dependence of $Y^{\rm min}$ on time reflects the fact, that entry into the loss-cone is a diffusive process}, and loss-cone depletion $t_{\rm out} \propto (Y^{\rm diff})^2$ \citep{ls77}, the contribution to $f(X,Y)$ of accretion should take into account the ratio
\begin{eqnarray}
q \equiv \frac{t_{\rm out}}{t_{\rm in}} \propto \frac{(Y^{\rm diff})^2}{(Y^{\rm min})^2}
\end{eqnarray}
\noindent
$Y^{\rm min} = J_{\rm z}^{\rm min}/ J_{\rm z}^{\rm max}$ and we denote the dimensionless angular momentum diffusion term due to gravitational scattering per time step.$(Y^{\rm diff})^2$ as $<(\Delta Y)^2>$

If $q < 1$, most loss-cone stars remain inside and $\Delta f(X,Y)$ is well represented by the flux of Eq.~\ref{yfluxdisc}, but, because stars could be scattered out of the loss-cone to orbits of $Y > Y_{\rm min}$ in an orbital time scale, a correction to the angular momentum flux is necessary. In the classical approximation, accretion of stars inside $Y_{\rm min}$ leads to an 'empty loss-cone' ($f(X,Y)=0$), but this is not a realistic boundary condition. If $q > 1$, the angular momentum diffusion term is larger than the loss-cone opening, so that most stars manage to scatter out of it and are not accreted. $q(E_{\rm crit})=1$, defines a critical energy ($E_{\rm crit}$) at a radius $r_{\rm crit}$, in the equatorial plane, which marks the transition between the 'full' and 'empty' loss-cone regimes.

This correction is implemented in the code by using a probability of accretion $P_a(X,Y,Y^{\rm diff})$, that a star do not escape from the loss-cone due to angular momentum change once it is inside, or it can fall into the loss-cone from outside. In the latter case, a probability of accretion $1-P_a$ is applied. The probability that an orbit with dimensionless angular momentum $Y$ suffers a change $\Delta Y < |Y-Y_{\rm min}|$, where $|Y-Y_{\rm min}|$ is the distance to the loss-cone boundary, is given by:
\begin{eqnarray}
P_a(Y) = \int_0^{\vert Y-Y^{\rm min} \vert} \frac{2}{\sqrt{\pi}Y^{\rm diff}} \exp{(-\frac{Y'^2}{(Y^{\rm diff})^2})} dY'
\label{prob}
\end{eqnarray}
i.e., centered at each ($X,Y$) grid cell a Gaussian distribution of orbits in $Y$ with the dispersion $Y^{\rm diff}$ is assumed.
Finally, the contribution of $f(X,Y)$ to accretion, at each energy-angular momentum grid cell is given by:
\begin{eqnarray}
{\Delta f}_{\rm acc} = P_a(Y) (f^{\rm old} + \Delta f )
\label{massdiff2}
\end{eqnarray}

$f^{\rm old}+ \Delta f$ is the result of redistribution of stars due to diffusion processes.  In order to get the total BH accretion mass, the distribution function $f(E,J_{\rm z})$ is integrated in real and phase space, as:

\begin{eqnarray*}
\Delta m_{\rm acc} = 4 \pi \int_0^{\varpi^{\rm tid}} \varpi d\varpi \int_0^{z^{\rm tid}} dz
\end{eqnarray*}
\begin{eqnarray}
\Bigl \lbrack \frac{2\pi}{\varpi} \int_{\phi_c}^{E_{\rm tid}} dE \int_{-J_{\rm z}^{\rm max}}^{+J_{\rm z}^{\rm max}} dJ_{\rm z} \ \Delta f_{\rm acc}(E,J_{\rm z)}  \Bigr \rbrack
\label{density}
\end{eqnarray}

\noindent
where $\varpi^{\rm tid}$ and $z^{\rm tid}$ give the tidal cluster radius in $\varpi$ and $z$ directions respectively. A factor of $2 \times 2 \pi = 4\pi$ is due to consideration of positive and negative zenithal coordinates and that the azimuthal component is symmetric. Moreover, $J_{\rm z}^{\rm max} = ± \varpi \sqrt{2(E-\phi)}$. The accretion mass is added to $M_{\rm bh} = M_{\rm bh}^{\rm old} + \Delta m_{\rm acc}$ and furthermore $M_{\rm cl} = M_{\rm cl}^{\rm old} - \Delta m_{\rm acc}$.
\section{Numerical results}
\label{sec:3}
\subsection{Initial conditions}
\label{sec:31}
\begin{figure}
\includegraphics[width=\linewidth]{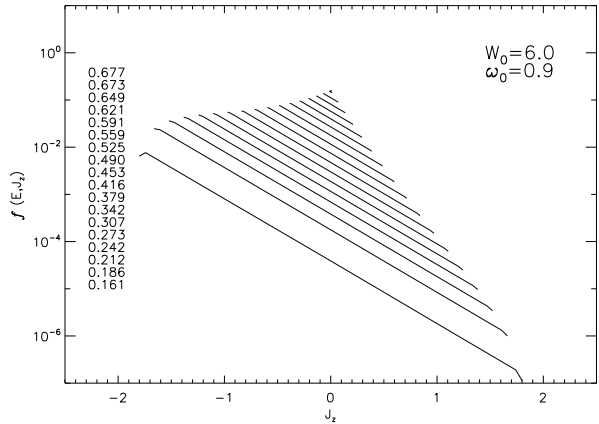}
\caption{$f(J_z)$, at constant energy $E$, for an initial model ($W_0=0.6$, $\omega_0=0.9$). From top to bottom the curves have smaller $|E|$, listed on the left column. Grid dimension is (200 $\times$ 201) in ($E,J_{\rm z}$)-space used to construct the models.}
\label{dfini}
\end{figure}

As initial configurations, truncated King models with added bulk motion are used. Their adopted distribution function is
\begin{equation*}
f(E,J_z) = \exp\left( -\frac{\Omega_0 J_z}{\sigma_{\rm c}^2} \right)\cdot  \left[ \exp (\frac{E_{\rm tid}-E}{\sigma_{\rm c}^2}) - 1 \right], \ E < E_{\rm tid}
\end{equation*}
\begin{equation}
f(E,J_z) = 0  \ \ \ \ \ \ \ \  \ \ \ \ \  \ \ \ \ \ \  \ \ \ \ \ \ \ \ \  \ \ \ \ \  \ \ \ \
\ \  \ \ \ \ \ \ \ \  \ \ \ \  \ \ \ \ \ \  \ \ \ \ \ \ ,  \ \ \ \ \ E > E_{\rm tid}
\label{king}
\end{equation}

\noindent
where $\sigma_{\rm c}$ is the central 1D velocity dispersion and $\Omega_0$ is an angular velocity. In Fig.~\ref{dfini}, $f(J_z)$ at constant energy $E$ against $J_{\rm z}$, for an initial high rotating model ($W_0=0.6$, $\omega_0=0.9$), is plotted. $f(J_z)$ covers a wide range of values in logarithmic scale, and  $J_{\rm z}$ varies from negative to positive values, accordingly to two directions of rotation around the z-axis. $J_{\rm z}=0$ represents stars on radial orbits, while $± J_{\rm z}^{\rm max}$, circular orbits. Note that the angular velocity $\Omega_0$ in Eq.~\ref{king} is given by the slope of $f$ in each curve of constant energy (a property of King models). The isoenergy sections become shorter, due to the smaller possible $J_{\rm z}^{\rm max}$ at higher absolute values of energy.

The system of units for the initial King models is given by
\begin{eqnarray}
G \equiv M_{\rm cli} \equiv r_{\rm ci} \equiv 1
\end{eqnarray}
where $M_{\rm cli}$ is the initial mass of the cluster and $r_{\rm ci}$ the initial core (King) radius
\begin{eqnarray}
r_{\rm c} \equiv \sqrt{\frac{9{\sigma_{\rm c}}^2}{4 \pi G n_{\rm c}}}
\end{eqnarray}
Where $n_{\rm c}$ is the central number density.

The initial conditions of each model are given by the pair ($W_0,\omega_0$), see Table ~\ref{initpar}. Here, $W_0$ is the familiar King parameter
\begin{eqnarray}
W_0 = \frac{(\phi(r_{\rm tid})-\phi_{\rm c})}{\sigma_{\rm c}^2} 
\end{eqnarray}
\noindent
where $\phi(r_{\rm tid})$ is the potential at the cluster tidal boundary and $\phi_{\rm c}$ the central potential.
\begin{eqnarray}
\omega_0= \sqrt{9/(4\pi G n_{\rm c})} \Omega_0
\end{eqnarray}
\noindent
is the initial rotational parameter. Radii are given in units of the initial cluster core radius. Table~\ref{initpar} shows initial parameters of the models. Intermediate models ($W_0=6.0$) are expected to reproduce current evolutionary states of most GCs. Their initial concentrations decrease and their dynamical ellipticities increase, the higher the initial rotation. $e_{\rm dyn}=1-b/a$ \footnote{axis ratio of an oblate spheroid} are calculated following \cite{goodman83} as defined in \cite{einsel99}. Tidally limited models presented in Section \ref{sec:5} are denoted by M1T to M5T.

\begin{figure}
\includegraphics[width=\linewidth]{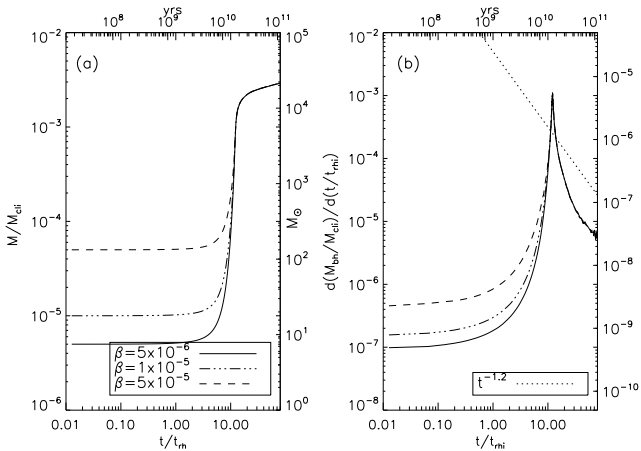}
\caption{\small Evolution of model M1 ($6.0,0.0$) for different values of $\beta$ ($\alpha = 2 \times 10^{-8}$). (a) BH mass against time. Left Y-axis shows code units ($M/M_{\rm cli}$) , right Y-axis transforms them to units of $M_\odot$ for a $M_{\rm cl_i} = 5 \times 10^6 M_\odot$. (b) Evolution of disruption rates. Left Y-axis shows code units ($d(M_{\rm bh}/M_{\rm cli})/d(t/t_{\rm rhi})$). Right Y-axis transforms them to units of $M_\odot/yr$.}
\label{timemassisocomp}
\end{figure}

There are two scaling parameters in the simulation: (i) the particle number N, which defines the mass of the single star to the total mass of the system, i.e. $m = M_{\rm cl_i}/N_{\rm i}$ (with $M_{\rm cl_i}=1$ in our units) and (ii) the mass of the black hole with respect to the total mass, which we specify in terms of initial seed black hole mass $\beta = M_{\rm bh_i} / M_{\rm cl}= 1 \times 10^{-5}$.
 Since we aim to study the dependence of the standard loss cone accretion model on the new physical situation of a rotating axisymmetric system surrounding the black hole, we fix $\beta$ and make our models scale-invariant to the particle number as long as the pure point-mass interactions are considered. The choice of $\beta$ is somewhat arbitrary, and it does not change much the physics, because Fokker-Planck models (e.g. \cite{amaro04}) show that the time evolution of the black hole and star cluster do not depend sensitively on the initial seed. In order to prove this statement, we performed tests with different values of $\beta$, as shown in Fig.~\ref{timemassisocomp}. All tests show a common final $M_{\rm bh}$ and the disruption rates follow the same self-similar evolution. For comparison, we include physical units in the right Y-axis and top X-axis. Here we applied our models to a massive globular cluster by using $M_{\rm i}= 5 \times 10^6 M_\odot$.
The initial $t_{\rm rh}$ was calculated following Eq.~\ref{trh}, after scaling the half-mass radius to the initial King radius using Table~\ref{initpar} ($r_{\rm h}=2.7 {\rm pc}$). In this equation, we use $N_{\rm i}= 5 \times 10^6$ and thus, $m = 1 M_\odot$. We obtain $t_{\rm rh} \sim 1.4 \times 10^9 yr$, which is a typical value of massive galactic globular clusters. In our models the initial seed represents technically a small perturbation in the central potential of our initial model, which accretes mass corresponding to the loss-cone accretion onto a fixed black hole, scaled down to our system (see below). This unphysical assumption for the initial accretion is used, because our goal is to study the long-term evolution of the system (which approaches a self-similar solution) and not the initial growth process.

\begin{figure}
\includegraphics[width=\linewidth]{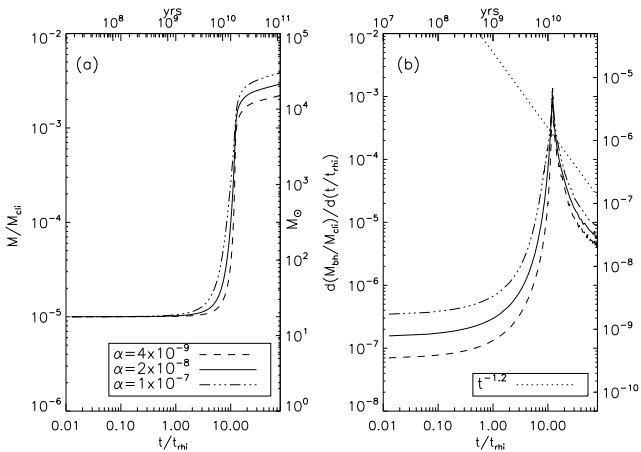}
\caption{\small Evolution of model M1 ($6.0,0.0$) for different values of $\alpha$ ($\beta = 1 \times 10^{-5}$ ). (a) BH mass against time.  (b) Evolution of disruption rate. Axis are labelled as in Fig.~\ref{timemassisocomp}}
\label{timemassisocompalpha}
\end{figure}

For stellar disruption we do have another parameter though, which is the stellar radius in our simulation units. This radius, defines then the disruption radius through Eq.~\ref{rdis} which grows in time due to black hole growth. We use as here for most simulations a value $\alpha = r_\star / r_{\rm cl} = 2 \cdot 10^{-8}$. In the following we use fiducial values of ($\alpha$, $\beta$) = ($ 2 \times 10^{-8},  1 \times 10^{-5}$) for most of our models, which (taking $r_\star$ as solar radius) define parameters for a globular cluster. We have tested the variation of $\alpha$, the results are shown in Fig.~\ref{timemassisocompalpha}. Changes of $\alpha$ by one order of magnitude influence our observables $M_{\rm bh}$ and $dM_{\rm bh}/dt$ only by a factor of 2-3. That suggests that our result can be applied by scaling to a wider range of astrophysical systems including galactic nuclei. Moreover, for a direct application to globular clusters our initial model is unphysical, because the seed black hole is not fixed, and it may grow or be ejected by close three-body encounters, all effects we are currently not taking into account. However, no matter what is the growth mechanism, if the black hole remains in the cluster and grows, we think that our scale-invariant solution should be reached. 

\begin{table}
  \centering
  \begin{tabular}{ | p{15pt} | c | c | c | c | c | c | c |} \hline
      Model &  $W_0$ & $\omega_0$ & $\ln{(\frac{r_{\rm tid}}{r_{\rm ci}})}$ & $\ln{(\frac{r_{\rm h}}{r_{\rm ci}})}$ & $e_{\rm dyn}$ & $t_{\rm rhi}$ & $\frac{T_{\rm rot}}{T_{\rm kin}}$\\  \hline \hline
    M0 & 3.0 & 0.0 & 5.79 & 1.50 &-0.001 & 29.40 & 0.00 \\ 
    M1 & 6.0 & 0.0 & 2.92 & 0.99 &-0.001 & 91.88 & 0.00 \\ 
    M2 & 6.0 & 0.3 & 2.71 & 0.96 & 0.105 & 87.73 & 7.00 \\ 
    M3 & 6.0 & 0.6 & 2.29 & 0.87 & 0.278 & 76.32 & 19.81 \\ 
    M4 & 6.0 & 0.9 & 1.92 & 0.83 & 0.403 & 71.24 & 30.25 \\ 
    M5 & 6.0 & 1.2 & 1.57 & 0.82 & 0.500 & 71.28 & 39.85 \\

 \hline
  \end{tabular}
   \caption[Initial model parameters]{\small Parameters of initial models used in the simulations. Column 1: model identification name;  Column 2: King potential; Column 3: dimensionless rotation; Column 4: concentration; Column 5: $\ln{(r_{\rm h}/r_{\rm ci})}$; Column 6: dynamical ellipticity; Column 7: initial half-mass relaxation time in code units; Column 8: rate of rotational energy to kinetic energy.}
  \label{initpar}
 \end{table}

\subsection{Numerical tests}
\label{sec:32}

We use a grid size of $N_X=200,N_Y=201,N_\varpi=N_z=200$ obtaining errors of angular momentum, mass and energy as shown in Table~\ref{error} for a typical model (our reference model, $W_0=6$) by the time the central density had decreased by about 2-3 orders of magnitude, during core expansion. As can be read from Table~\ref{error} a grid convergence study shows that shorter grids are not sufficient to keep the errors small and in order to achieve and improve the accuracy reported by Fokker-Planck calculations without deep potentials of 1.7, 0.7 and 0.4 per cent in energy, mass and angular momentum respectively, \citep{einsel99}.

\begin{table}
  \centering
  \begin{tabular}{ | c | c | c | c |} \hline
      Grid &  $\Delta E/E$ & $\Delta M/M$ & $\Delta J_z/J_z$ \\  \hline \hline
    50x50 & $7.85 \cdot 10^{-2}$ & $3.17 \cdot 10^{-2}$ & $3.66 \cdot 10^{-2}$ \\ 
    100x100 & $1.84 \cdot 10^{-2}$ & $8.43 \cdot 10^{-3}$ & $5.88 \cdot 10^{-3}$ \\
    200x200 & $2.08 \cdot 10^{-3}$ & $2.89 \cdot 10^{-4}$ & $8.66 \cdot 10^{-4}$\\
  \hline
  \end{tabular}
   \caption[Relative errors]{\small Convergence analysis of relative errors performed by the code. Column 1: Grid size;  Column 2: Relative energy error; Column 3: Relative mass error; Column 4: Relative angular momentum error}
  \label{error}
 \end{table}

There are mainly two effects that make our numerical problem much more difficult than earlier ones, increasing the errors. First, due to the axisymmetry of the system; and secondly, due to the deep growing central potential. It made necessary a better resolution in real and velocity space, and specially at the time the region of influence of the BH dominates the core and during expansion. We increased the resolution in the inner parts of the spatial grid ($N_\varpi,N_z$) by setting a linear grid for the core and a logarithmic one for the outer regions. The ($N_X,N_Y$)-grid has correspondingly a higher resolution for absolute energies larger than $E_{\rm crit}$, as mentioned in Section \ref{sec:21}. We set here $E_0 = E_{\rm crit}$ in Eq.~\ref{xgrid}. 
 A further test of the code was made by reproducing the spherically symmetric case with the non-rotating initial parameters (as presented in Section \ref{sec:41}). Comparison to NBody realizations are being presented in a forthcoming paper. But see \cite{kim08} for a comparative study of non-BH systems.

\begin{figure}
\includegraphics[width=\linewidth]{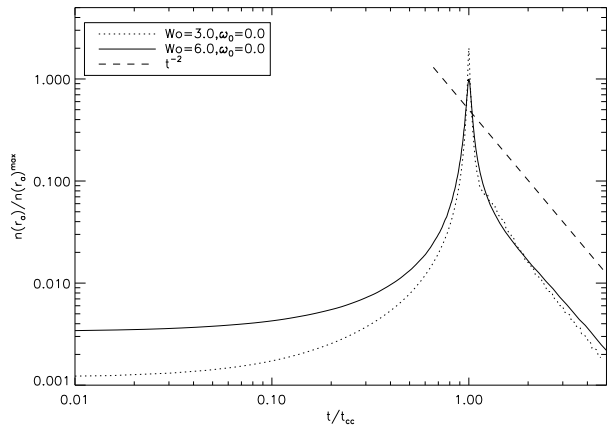}
\caption{\small Evolution of density at the influence radius for models of different initial concentration ($W_0 = 3.0, 6.0$). The comparable self-similar expansion after collapse shows that the evolution in relaxation time scales is not sensitive to the initial configuration of the system. Density of model $W_0=6.0$ was normalized to its density maximum and to its collapse time. Model $W_0=3.0$ was overplotted for comparison reasons.}
\label{timerhoisocomp}
\end{figure}

As is well known, the long-term evolution of relaxed systems does not depend on the details of the initial conditions, as these are erased on a relaxation time-scale. To verify this we have also started some of our runs from an initial King profile, with a concentration parameter $W_0=3.0$ (Model M0). The evolution of the density at the influence radius is compared to the model M1 ($W_0=6.0$) in Fig.~\ref{timerhoisocomp}. Both the  $W_0= 3$  and the $ W_0= 6$  models experience a self-similar expansion, which approximates $ n_{\rm a} \sim t^{-2}$ after collapse is prevented. From now on we define the collapse time as the time of maximum density at the influence radius in the system (see Tables~\ref{colltiso} and \ref{collttid}). In  Fig.~\ref{timerhoisocomp} the evolution of model $W_0=6.0$ was normalized to its density maximum and to the collapse time ($t_{\rm cc}$), while the model $W_0=3.0$ was overplotted for comparison reasons.

Typical runs for the evolution of one model up to $\sim 50 \ t_{\rm rhi}$ needed about 40 hours on a 3-GHz Pentium IV processor (ARI-ZAH, University of Heidelberg). A speed-up through parallel processing would be recommended for multi-mass versions of the present code (work in progress). This performance is not disappointing, taking into account that the number of floating-point operations performed per time-step in our models is $N_X \times N_Y \times N_\varpi \times N_z$ and since the time steps get much shorter close to $t_{\rm cc}$. The results presented here concentrate on our standard model $W_0=6$.

\section{Isolated systems}
\label{sec:4}
\subsection{Spherical symmetry}
\label{sec:41}
In order to test the method, we reproduce the evolution of isolated dense stellar systems in the spherically symmetric case. We realize this model by setting the initial rotating parameter $\omega_0$ to zero.
$M_{\rm bh_i}$ starts growing through accretion of stars in low-$J_{\rm z}$ orbits. For $\sim 0.3 t_{\rm rhi}$ the evolution is unaffected by the presence of the small central BH and is dominated by the contraction of the core. But the increasing density supports the growth rate of $M_{\rm bh}$ in later times, when the BH-potential ($ \sim GM_{\rm bh}/r$) dominates the stellar distribution within $r_{\rm a}$. 

\begin{figure}
\includegraphics[width=\linewidth]{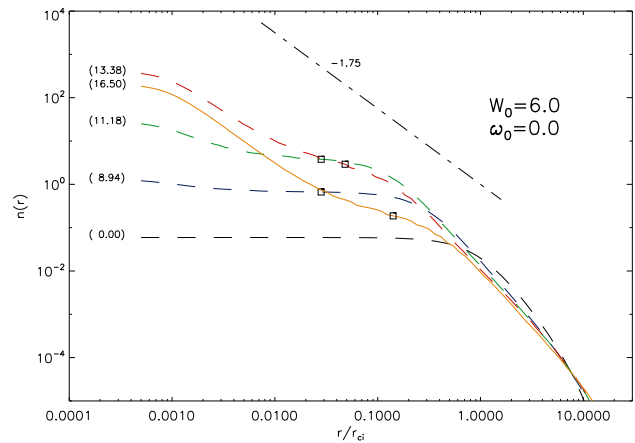}
\caption{\small Equatorial density profile ($z=0$) for model M1 at different times (in parenthesis) given in units of initial half-mass relaxation time ($t_{\rm rhi}$). Dashed colored lines (black, blue, green and red) show the evolutionary profiles and the orange solid line the final profile. The dot-dashed line shows the -7/4 slope. The location of $r_{\rm a}$ is shown as squares.}
\label{timerho060000iso}
\end{figure}

The final steady-state, solid curve in Fig.~\ref{timerho060000iso}, evolves towards a power-law of $\lambda =-1.75$, according to $n \propto r^{\lambda}$. This solution has been extensively studied in the spherical case by \cite{bah76,ls77,mar80} and others. It forms inside $r_{\rm a}$ and is maintained in the post-collapse phase, while the evolution is driven through energy input from the central object, dominating always larger zones. The density profile flattens close to the center due to the effective loss-cone accretion and it remains practically unchanged in the halo, where the loss-cone loses its significance.

\begin{figure}
\includegraphics[width=\linewidth]{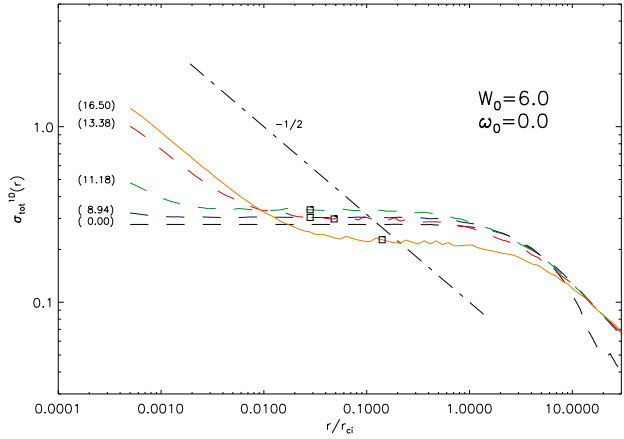}
\caption{\small Equatorial profile ($z=0$) of 1D total velocity dispersion as in Fig.~\ref{timerho060000iso} for the density (Model M1). The dot-dashed line shows the -1/2 slope and the dashed lines the evolutionary profiles (as in Fig.~\ref{timerho060000iso}). The location of $r_{\rm a}$ is shown as squares. The curve achieves the steady state (orange solid line) in $\sim 10 t_{\rm rhi}$}
\label{timevdisp060000iso}
\end{figure}

As the systems evolves, orbits in the region of influence of the BH become Keplerian bounded. Their velocity dispersion approximates a power-law of -1/2 within the BH influence radius $r_{\rm a}$. Fig.~\ref{timevdisp060000iso} shows the evolution of the total one-dimensional velocity dispersion profile in the same way as Fig.~\ref{timerho060000iso} does for the density. The velocity dispersion grows significantly inside $r_{\rm a}$ and faster when the cluster is close to collapse, due to the presence of the deep central potential.

\begin{figure}
\includegraphics[width=\linewidth]{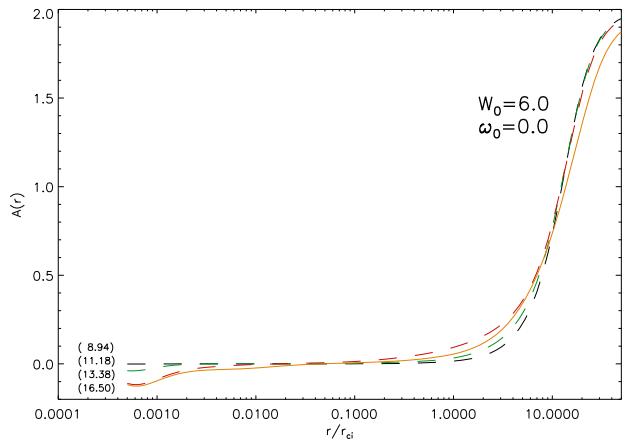}
\caption{\small Anisotropy, defined as $A \equiv 2(1-\frac{{\sigma_\phi}^2}{{\sigma_{\rm r}}^2})$, against radius for the same model of Figs.~\ref{timerho060000iso} and \ref{timevdisp060000iso}. Small tangential anisotropy forms inside $r_{\rm a}$. Radial anisotropy dominates the halo. Curves are labeled as in Figs.~\ref{timerho060000iso} and \ref{timevdisp060000iso}}
\label{timeaniso060000iso}
\end{figure}

Fig.~\ref{timeaniso060000iso} shows the anisotropy profile in the system at different times. Anisotropy is defined as $A \equiv 2(1-\frac{{\sigma_\phi}^2}{{\sigma_{\rm r}}^2})$, where  the total velocity dispersion $\sigma_t^2=\sigma_\phi^2+2\sigma_r^2$ ($\sigma_r=\sigma_z$). Velocity dispersion $\sigma_\phi$, which is the azimuthal velocity dispersion in the direction of rotation and  $\sigma_r$ are calculated initially by taking moments of $f$ with respect to $\vec{v}$ and $\vec{v}^2$ and assuring conservation of energy and angular momentum per unit volume by encounters between stars (\cite{goodman83}). The initial profile shows a maximum positive halo anisotropy (radial orbits dominate the halo in the initial configuration) after a very short time (a fraction of $t_{\rm rh_i}$). The total amount of radial anisotropy, by using the rate $K_{\rm r}/K_{\phi}$, where $K_r$ is the kinetic energy in the radial degree of freedom and $K_\phi$ is the kinetic energy in the tangential degree of freedom, gives a maximum excess of 13 \% in  $K_r$ present in the system at the time of the final profile showed in Fig.~\ref{timeaniso060000iso}. The small excess of $K_{\rm r}$ can be understood, because $A(r)$ rises only in the outer regions of the system where the density is low. This results are of the same order like the presented in previous theoretical studies of anisotropy profiles and evolution of globular clusters \citep{louis91,giersz94}. Moreover, small negative anisotropy forms slowly inside the BH influence radius (tangential orbits dominate the center close to the BH), while radial anisotropy remains in the halo \citep{quinlan95,freitag02,baum04}.

\begin{figure}
\includegraphics[width=\linewidth]{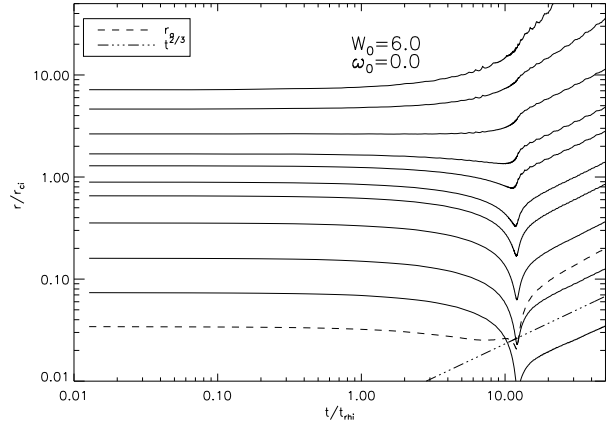}
\caption{\small Evolution of Lagrangian radii $r_{55}$, as defined in text, for the model M1. The time is given in units of $t_{\rm rhi}$. Solid lines represent from bottom to top radii of 0.01~\%, 0.1~\%, 1~\%, 5~\%, 10~\%, 20~\%, 30~\%, 50~\%, 75~\%, 90~\% of the total mass of the system. The influence radius is additionally plotted (dashed line) and the self-consistent expansion is shown by the dot-dashed line $r \propto t^{2/3}$.}
\label{timelagr060000iso}
\end{figure}

Evolution of Lagrangian radii is a good indicator for the contraction and further reexpansion of mass shells. During expansion core shells increase as r $\propto t^{2/3}$ , as expected for a system in which the central object has a small mass and the energy production is confined to a small central volume \citep{henon65,shapiro77,mcmillan81,goodman84}. In Fig.~\ref{timelagr060000iso} the influence radius is plotted additionally to the Lagrangian radii. $r_{55}$ refers to the evaluation of Lagrangian radii at a zenithal angle, where the effects of probable flattening on the mass columns are expected to be less important, that deviations from spherical symmetry are only up to second order in a Legrende expansion, i.e. $P_2(cos \theta)=0$. That gives $\theta=54,74$ \citep{einsel99}.

Figs.\ref{timemassisocomp} and \ref{timemassisocompalpha} show how $M_{\rm bh}$ reaches a nearly constant fraction of $M_{\rm cli}$ at collapse time ($t_{\rm cc}$), while the star accretion rate ($dM/dt$) is maximal at $t_{\rm cc}$ due to the higher density of orbits in the core decreasing afterwards very rapidly (Figs.~\ref{timemassisocomp}b and \ref{timemassisocompalpha}b). For a density power-law of $\lambda =-1.75$, the expected proportionality $dM/dt \propto t^\alpha$ turns out to be $\alpha = -1.2$ \citep{amaro04}.

During evolution, the core is heated via the consumption of stars in bound, high energetic orbits in the cusp. Energy flux is achieved by small-angle, two-body encounters, by which some stars lose energy and move closer to the BH being eventually consumed, while the stars with which they interact gain energy and move outward from the cusp into the ambient core. Angular momentum transport is initially enhanced by gravo-gyro instabilities and not affected by BH accretion of stars on orbits of low $J_{\rm z}$. Later, when core density grows, mass growth rate increases strongly due to core contraction to higher densities and stronger stellar interaction.

The general behavior confirms previous studies of spherically symmetric systems. \cite{mar80} follow the evolution of a star cluster containing a central BH (included in their simulations at collapse time). The BH mass stalls after approximately 2 relaxation time units to a final mass of $\sim 4000 M_\odot$. In our models, a similar rapid evolution before expansion is observed, and the final masses are comparable (see Tables~\ref{massiso} and \ref{masstid}).

\subsection{Axisymmetric isolated systems}
\label{sec:42}

\begin{figure}
\includegraphics[width=\linewidth]{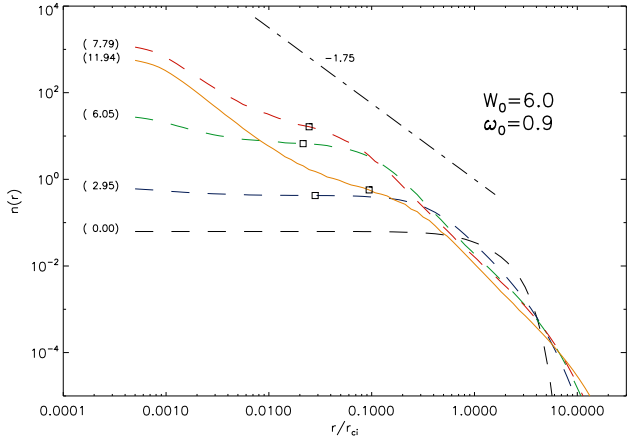}
\caption{\small Equatorial density profile ($z=0$) for model M4 ($W_0=6.0$,$\omega_0=0.9$) for different times given in units of $t_{\rm rhi}$. The orange solid line shows the final profiles and the dashed lines the evolutionary profiles. The dot-dashed line shows the -7/4 slope. Colors are like in Fig.~\ref{timerho060000iso}. The location of $r_{\rm a}$ is shown as squares.}
\label{timerho060090iso}
\end{figure}

Evolution of the density profile of model M4 is shown in Fig.~\ref{timerho060090iso}. The extent of the BH gravitational influence is marked on each curve at the position of $r_{\rm a}$ (squares). Evolutionary profiles are represented by dashed curves. Note that the limit between the cusp and the core is located at $\sim r_{\rm a}$. The central cusp in the density profile grows first very slow and faster towards core collapse. It approaches the -7/4 cusp, like in the spherically symmetric case.

\begin{figure}
\includegraphics[width=\linewidth]{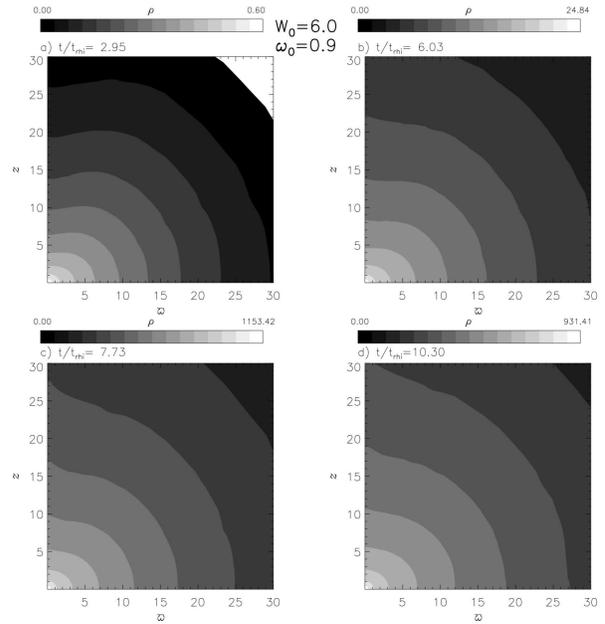}
\caption{\small Evolution of density distribution in the meridional plane for a model M4. Cylindrical coordinates ($\varpi$, $z$) are used. Lighter zones represent higher isodensity contours. The time is given in units of initial half-mass relaxation time ($t_{\rm rhi}$).}
\label{rhomtime060090iso}
\end{figure}

\begin{figure}
\includegraphics[width=\linewidth]{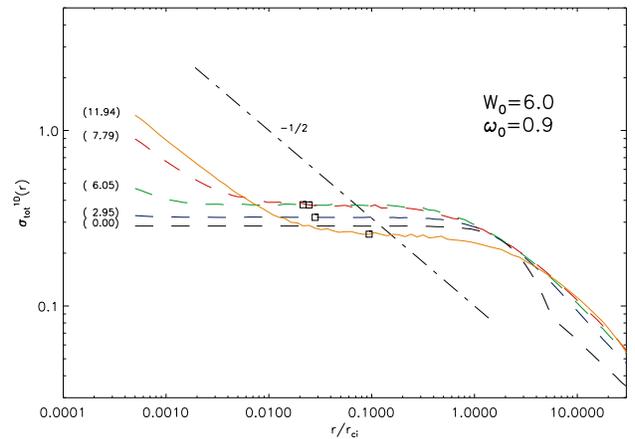}
\caption{\small Equatorial profile ($z=0$) of 1D total velocity dispersion as in Fig.~\ref{timerho060090iso} for the density (Model M4). The dot-dashed line shows the -1/2 slope and the dashed lines the evolutionary profiles. The location of $r_{\rm a}$ is shown as squares.}
\label{timevdisp060090iso}
\end{figure}

Fig.~\ref{rhomtime060090iso} shows the evolution of density contours in the meridional plane ($\varpi,z$). In the regions where BH star accretion dominates (i.e. inside $r_{\rm a}$), the isodensity contours grow stronger due to the presence of the BH (lighter zones in Fig.~\ref{rhomtime060090iso}). The cusp forms a strong gradient towards the center, forming very fast close to core collapse. Note that the flattened shape of the system remains at later times, during the expansion phase (Fig.~\ref{rhomtime060090iso}d). The velocity dispersion is Keplerian within the BH influence radius $r_{\rm a}$, as Fig.~\ref{timevdisp060090iso} shows. Extension of $r_{\rm a}$ is comparable to non-rotating models but cusp formation time is shorter, the higher the initial rotation parameter (from comparison to Fig.\ref{timevdisp060000iso}) due to the faster evolution of this models. Moreover, the rate $K_{\rm r}/K_{\rm t}$ is larger the higher the rotation, with maximum values of 1.18, 1.21, 1.23 and 1.26 for the models M2 to M5, respectively.

\begin{figure}
\includegraphics[width=\linewidth]{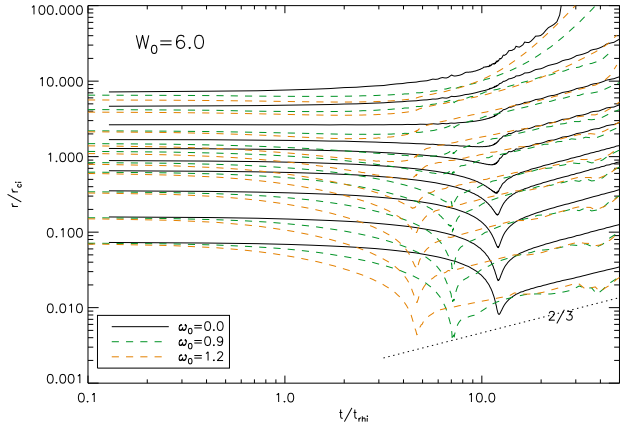}
\caption{\small  Evolution of mass shells (Lagrangian radii $r_{55}$) for the models M1, M4 and M5. Solid lines represent from bottom to top radii of 0.01~\%, 0.1~\%, 1~\%, 5~\%, 10~\%, 20~\%, 30~\%, 50~\%, 75~\%, 90~\% of the total mass. Self-similar expansion is shown by the dotted line. The time is given in units of $t_{\rm rhi}$.}
\label{timelagr060iso}
\end{figure}

Evolution of Lagrangian radii containing the indicated fractions of the initial mass is shown in Fig.~\ref{timelagr060iso} in comparison to the non-rotating model. Lagrangian radii give also a qualitative description of the interaction of a growing BH and the cluster mass shells. Initially, the BH mass growth is slow due to the low central density, and Lagrangian radii are dominated by core contraction. Finally the collapse is halted and reversed and the mass shells re-expand. It happens faster for higher rotating models. Note that the smallest radius contains only 0.01 \% of the cluster mass, in order to follow the evolution of mass shells closer to the BH. Our single mass rotating models show some deviations of the self-similar expansion phase, which should be further investigated.

\begin{figure}
\includegraphics[width=\linewidth]{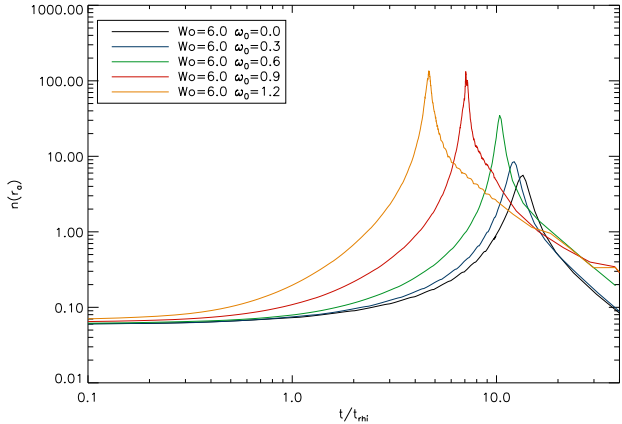}
\caption{\small Evolution of density at $r_{\rm a}$. Density maxima from right to left correspond to the higher rotational parameter $\omega$. The time is given in units of $t_{\rm rhi}$.}
\label{timerho060iso}
\end{figure}

Fig.~\ref{timerho060iso} shows the evolution of density at $r_{\rm a}$ for models with rotation parameters $\omega_0$=0.0 (non-rotating), 0.3, 0.6, 0.9 and 1.2. After a similar initial evolution, collapse is faster the higher the initial rotation. Gravothermal and gravo-gyro instabilities drive collapse, while angular momentum is transported out of the core in a progressively more efficient way for the higher rotating models. Both instabilities occur together and support each other. The collapse is reversed due to the energy source built by the star accreting BH, while the central density drops during expansion. Because radial anisotropy first dominates (as shown in Fig.~\ref{timeaniso060000iso}), it supports BH accretion of stars in the core, which are able to interact with these low-$J_{\rm z}$ (eccentric) orbits in the outer parts.

\begin{table}\centering
\begin{tabular}{ | c || c | c |} \hline 
Model & $\frac{t_{\rm cc}}{t_{\rm rhi}}$ non-BH models & $\frac{t_{\rm cc}}{t_{\rm rhi}}$ BH models\\
\hline \hline 
M1 &  12.36 & 12.20  \\ 
M2 &  11.63 & 11.40  \\ 
M3 &  9.51 &  10.44  \\ 
M4 &  6.95 &   7.07  \\
M5 &  4.71 &   4.60  \\
\hline
\end{tabular}\\
\caption[Collapse times]{Comparison of collapse times  $t_{\rm cc}$ between rotating BH and non-BH models}
\label{colltiso}
\end{table}     

As seen in Table~\ref{colltiso} collapse times for non-BH models are comparable to the BH rotating models and are as well shorter for higher initial rotation. $t_{\rm cc}$  varies from $12.20 t_{\rm rh}$ (non-rotating models) to $4.6 t_{\rm rh}$ (high rotating models). At $t_{\rm cc}$ angular momentum diffusion is more effective, due to the interplay between dynamical instabilities and BH star accretion.
\begin{table}
  \centering
  \begin{tabular}{ | c | c | c | c | c |} \hline
      Model&$\frac{M_{\rm bh}^{\rm stall}}{M_{\rm cli}}$&$M_{\rm bh}^{\rm stall} (M_\odot)$&$\frac{d(M/M_{\rm cli})}{d(t/t_{\rm rhi})}_{\rm max}$&$\frac{dM}{dt}_{\rm max} (\frac{M_\odot}{\rm yr})$\\
        \hline \hline
    M1 & $3.0 \cdot 10^{-3}$ & $1.5 \cdot 10^4$ &$1.00 \cdot 10^{-3}$ & $5.0 \cdot 10^{-6}$\\ 
    M2 & $2.4 \cdot 10^{-3}$ & $1.2 \cdot 10^4$ &$1.05 \cdot 10^{-3}$ & $5.3 \cdot 10^{-6}$\\ 
    M3 & $2.0 \cdot 10^{-3}$ & $1.0 \cdot 10^4$ &$1.24 \cdot 10^{-3}$ & $6.2 \cdot 10^{-6}$\\ 
    M4 & $1.7 \cdot 10^{-3}$ & $8.5 \cdot 10^3$ &$1.68 \cdot 10^{-3}$ & $8.4 \cdot 10^{-6}$\\ 
    M5 & $1.5 \cdot 10^{-3}$ & $7.5 \cdot 10^3$ &$1.57 \cdot 10^{-3}$ & $7.9 \cdot 10^{-6}$\\ \hline
  \end{tabular}
   \caption[Evolution of mass parameters ]{\small Evolution of mass parameters. Column 1: Model name; Column 2: Final $M_{\rm bh}$ in units of initial mass as it reaches an asymptotic mass; Column 3:  $M_{\rm bh}$ in $M_\odot$-units; Column 4: maximal accretion rate $dM/dt$ in units of initial mass and half-mass relaxation time; Column4: $dM/dt$ in $(M_\odot$/yr)-units.}
  \label{massiso}
 \end{table}

Table~\ref{massiso} shows the final BH mass (in units of $M_\odot$) of each model and the respective maximal accretion rates (in solar mass per year) at the time $M_{\rm bh}$ stalls and the accretion rate begins to slowdown. For a system of $M_{\rm cl_i} = 5 \cdot 10^6 M_\odot$, $M_{\rm bh}^{\rm stall}$ varies between $7.5 \cdot 10^3 M_\odot$ and $1.5 \cdot 10^4 M_\odot$, which agrees with the IMBH estimated by theoretical studies and observations of globular clusters \citep{geb00,geb02a,ger02}, as expected according to the initial conditions of our models (Sect.~\ref{sec:31}). Physical units were derivated as described in \ref{sec:31} using the initial parameters of the corresponding King models (Table~\ref{initpar}). The general behavior exhibits a slightly decreasing $M_{\rm bh}^{\rm stall}$ but higher mass growth rates for higher initial rotation. In rotating models, a stalling of $M_{\rm bh}$ happens always faster the higher $\omega_0$ is (Table~\ref{colltiso})

\begin{figure}
\includegraphics[width=\linewidth]{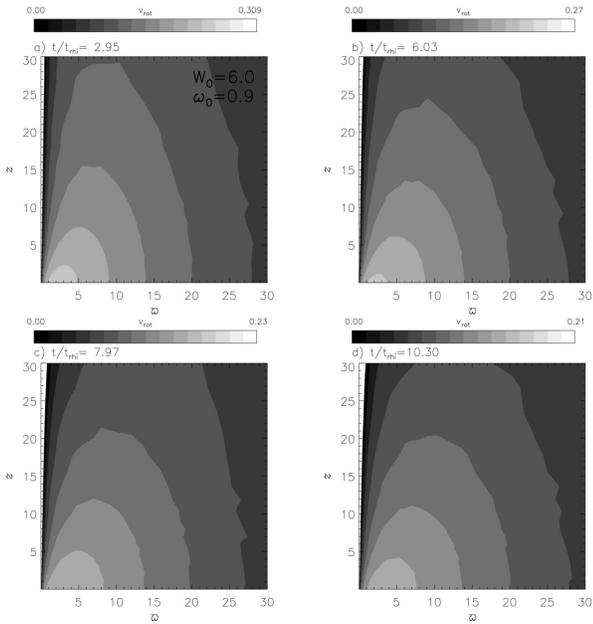}
\caption{\small 2D rotational velocity distribution in the meridional plane for the model M4 ($6.0,0.9$).}
\label{vrotmtime060090iso}
\end{figure}

It is known, that in rotating models without BH, the total collapse time is shortened by the gravo-gyro effect \citep{hachisu79,hachisu82}, by which large amounts of initial rotation drive the system into a phase of strong mass-loss while it contracts (the core rotates faster although angular momentum is transported outwards). At the same time, the core is heating, while the source of the so-called 'gravo-gyro' catastrophe is consumed and the growth in central rotation levels off after 2 - 3 $t_{\rm rh}$ towards core collapse \citep{einsel99}. Simulations into post-collapse phase, driven by three-body binary heating shown by \cite{kim02} exhibit a faster evolution for rotating models.
BH rotating models experience in a similar way, the onset of gravo-gyro instabilities, as angular momentum diffuses outwards, leading to an increase of central rotation \citep{hachisu79,hachisu82}. Moreover, BH mass growing causes the expansion of the system (Fig.~\ref{timelagr060000iso} and Fig.~\ref{timelagr060iso}), and leads to an ordered motion of high-$J_{\rm z}$ bounded orbits around the central BH (tangential anisotropy) supporting development of central rotation. Nonetheless, as the BH reaches its final mass, angular momentum continues being transported out of the core.
Fig.~\ref{vrotmtime060090iso} shows snapshots of the evolution of the 2-dimensional distribution of $v_{\rm rot}$ in the meridional plane, at representative times, where the lighter areas represent contours of higher rotation. Note that an important amount of central rotation is still present during the time of expansion Fig.~\ref{vrotmtime060090iso}d (and compare to Fig.~\ref{vrotmtime060090tid}).

\begin{figure}
\includegraphics[width=\linewidth]{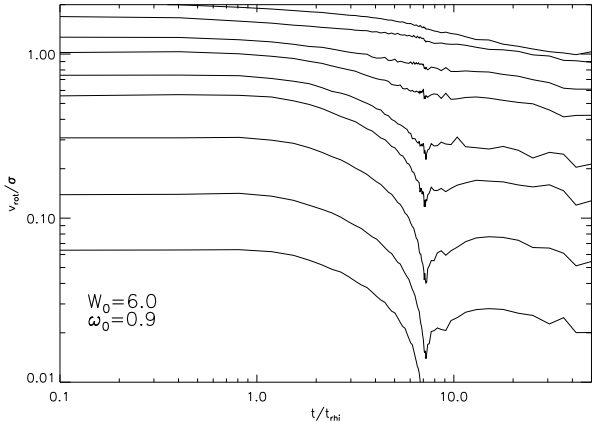}
\caption{\small  $V_{\rm rot} / \sigma$ at the Lagrangian radii as a function of time for the model M4 ($6.0,0.9$).  $V_{\rm rot} / \sigma$ at radii of 0.01~\%, 0.1~\%, 1~\%, 5~\%, 10~\%, 20~\%, 30~\%, 50~\%, 75~\% of the total mass are represented by solid lines from bottom to top. The time is given in units of $t_{\rm rhi}$.} 
\label{timevrotsiglag060090iso}
\end{figure}

The rate of rotational velocity over velocity dispersion represents the importance of ordered motion in comparison to random motion.
In Fig.~\ref{timevrotsiglag060090iso} the evolution of the rate $V_{\rm rot} / \sigma$ at the Lagrangian radii is shown. Up to collapse, there is no considerable influence of the BH, while during the expansion $V_{\rm rot} / \sigma$ grows slightly in time, specially for the inner shells.

The results presented here support the thesis that the formation of a massive central dark object, could predict the remaining of central rotation in GCs over long evolutionary time scales. Nonetheless, the central $V_{\rm rot} / \sigma$ finally decreases due to the growing central velocity dispersion, and falls later after $t_{\rm cc}$ together with $V_{\rm rot}$, while angular momentum is carried away from the system. In the outer regions the effect is smaller, maintaining a slower decreasing.

\section{Tidally limited models}
\label{sec:5}
\begin{figure}
\includegraphics[width=\linewidth]{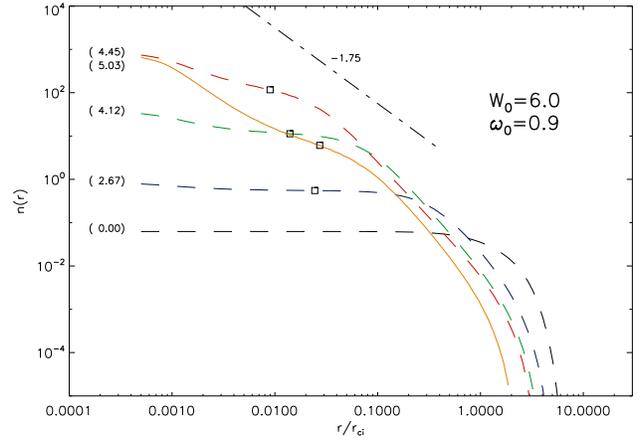}
\caption{\small Equatorial density profile ($z=0$) for model M4T (tidally limited case). Times are given in parenthesis and in units of $t_{\rm rhi}$. The dot-dashed line shows the -7/4 slope and the dashed lines the evolutionary profiles. Solid (orange) line shows the final profile. The location of $r_{\rm a}$ is shown as squares.}
\label{timerho060090tid}
\end{figure}

In this models, mass-loss is included, allowing the escape of stars through the energy tidal limit (see Fig.~\ref{losscone}). While $M_{\rm bh}$ grows and central density increases within $r_{\rm a}$, the system loses mass through the outer tidal boundary due to relaxation effects. Evolution of the density profile of model M4T is shown in Fig.~\ref{timerho060090tid}. The extent of the BH gravitational influence is marked on each curve at the position of the influence radii $r_{\rm a}$ (squares). Evolutionary profiles are represented by dashed curves. The power-law of $\lambda =-1.75$ is not completely reached as in the isolated case (Fig.~\ref{timerho060090tid}), probably due to the strong mass loss. In the halo, it gets steeper beyond $r_{\rm a}$ up to the tidal radius ($r_{\rm tid}$), where the loss-cone loses its significance. $r_{\rm tid}$ itself becomes smaller in time as a consequence of tidal mass-loss.

\begin{figure}
\includegraphics[width=\linewidth]{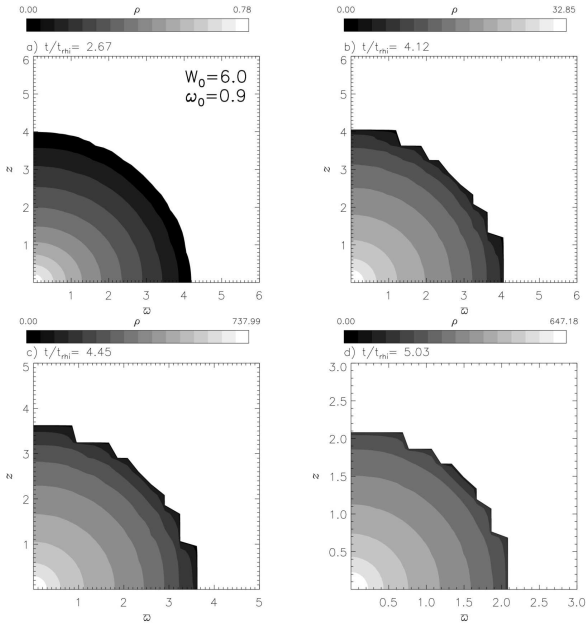}
\caption{\small Evolution of density distribution in the meridional plane for a model M4T in the tidally limited case. Cylindrical coordinates ($\varpi$, $z$) are used. Lighter zones represent higher isodensity contours. Note that scales are different in the bottom figures due to the shrinking of the outer tidal radius.}
\label{rhomtime060090tid}
\end{figure}

\begin{figure}
\includegraphics[width=\linewidth]{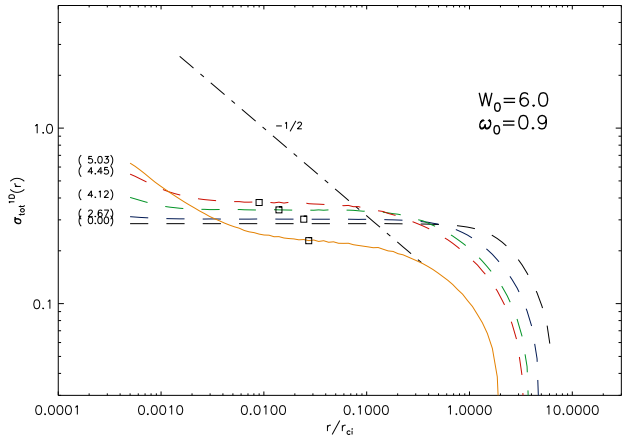}
\caption{\small Equatorial profile ($z=0$) of 1D total velocity dispersion (Model M4T) in the tidally limited case. The dot-dashed line shows the -1/2 slope. Evolutionary profiles are labeled like in Fig.~\ref{timerho060090tid}. The location of $r_{\rm a}$ is shown as squares.}
\label{timevdisp060090tid}
\end{figure}

Fig.~\ref{rhomtime060090tid} shows the evolution of density in the meridional plane ($\varpi,z$). In the regions where BH star accretion dominates (i.e. inside $r_{\rm a}$), the isodensity contours grow due to the presence of the BH (lighter zones in Fig.~\ref{rhomtime060090tid}). Note that scales are different for the bottom figures due to the shrinking of the system which loses mass through the outer tidal boundary, and the cluster tidal radius becomes smaller (darker areas). The shape of the system becomes faster more spherical that in the isolated case. Like in the isolated case, orbits in the region of influence of the BH become Keplerian bounded. Their velocity dispersion grows towards a power-law of -1/2 within the BH influence radius $r_{\rm a}$ (Fig.~\ref{timevdisp060090tid}).

\begin{figure}
\includegraphics[width=\linewidth]{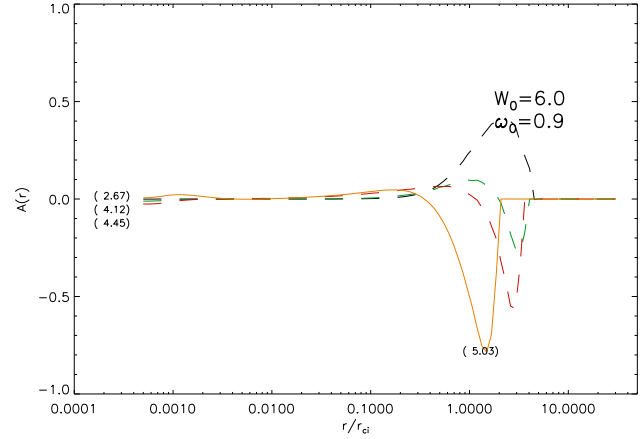}
\caption{\small Anisotropy, defined as $A \equiv 2(1-\frac{{\sigma_\phi}^2}{{\sigma_{\rm r}}^2})$, against radius for the same model of Figs.~\ref{timerho060090tid} and \ref{timevdisp060090tid}. Dashed curves are evolutionary profiles (labeled with times in parenthesis). Solid (orange) curve (latest evolutionary time) is labeled separately.}
\label{timeaniso060090tid}
\end{figure}

\begin{figure}
\includegraphics[width=\linewidth]{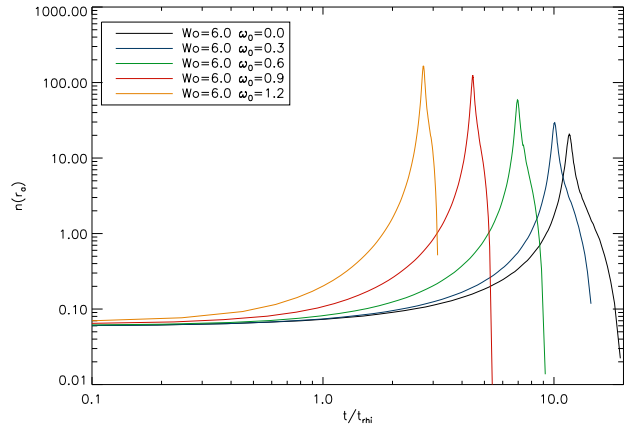}
\caption{\small Evolution of density at the influence radius for models of $W_0=6.0$ and initial rotation $\omega_0=$0.0,0.3,0.6,0.9,1.2 (density maxima from right to left).}
\label{timerho060tid}
\end{figure}

Fig.~\ref{timeaniso060090tid} shows the anisotropy profile in the system at different times. The initial profile shows radial halo anisotropy (radial orbits dominate the halo in the initial configuration). Tangential anisotropy seems not to form inside the BH influence radius, as it does in the isolated case. Moreover, at later times, tangential orbits dominate the halo as a consequence of an effective $J_{\rm z}$-transport outwards and the accretion of preferentially radial orbits by the central BH. In general, a faster evolution in higher rotating models leads to a smaller $M_{\rm bh}^{\rm stall}$ (see Table~\ref{masstid}), and thus to a smaller tangential anisotropy (i.e. smaller in model M4T than M1T). This was expected, since a more massive BH will consume more stars in preferentially radial orbits and the life time of high rotating tidally limited systems is too short to develop higher $M_{\rm bh}$. Note that at this evolutionary times ($t \sim 5 t_{\rm rhi}$ for M4T) the cluster has lost more than 50 \% of its mass (see Table~\ref{collttid}) and the densities in the outer parts are much lower than in the isolated models. As a consequence, the measured rates $K_{\rm r}/K_\phi$ range from 0.98 to almost 1.00 for all these models. At the same time, $r_{\rm a}$ becomes larger and dominates almost the hole system, which itself is close to dissolution (see Fig.~\ref{timerho060tid})

Fig.~\ref{timerho060tid} shows the evolution of central density for all models, with rotation parameters $\omega_0$=0.0 (non-rotating), 0.3, 0.6, 0.9 and 1.2 for the tidal limited case. Collapse time is reached faster, the higher the initial rotation, in comparison to the isolated models. Gravothermal and gravo-gyro instabilities support each other and collapse phase is reversed due to the energy source built by the star accreting BH.

\begin{figure}
\includegraphics[width=\linewidth]{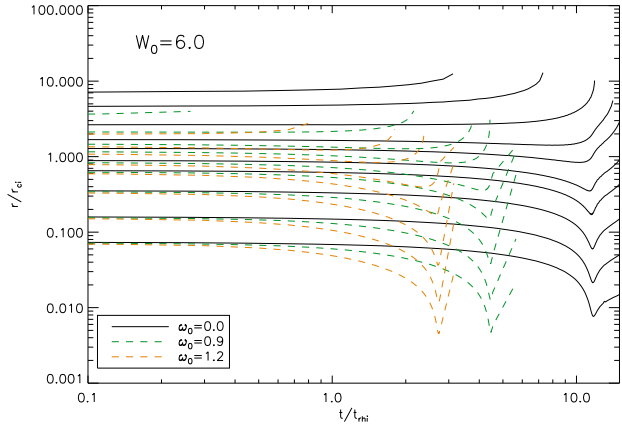}
\caption{\small Evolution of mass shells (Lagrangian radii $r_{55}$) for the models M1T, M4T and M5T.  Solid lines represent from bottom to top radii of 0.01~\%, 0.1~\%, 1~\%, 5~\%, 10~\%, 20~\%, 30~\%, 50~\%, 75~\%, 90~\% of the total mass. Note that some of the outer Lagrangian radii in the rotating models disappear at very early times, due to strong mass loss.}
\label{timelagr060tid}
\end{figure}

Evolution of Lagrangian radii containing the indicated fractions of the initial mass is shown in Fig.~\ref{timelagr060tid}. Due to mass-loss, the outer mass shells are rapidly truncated, the faster the higher initial rotation. At the same time, the core density grows (higher disruption rates). Later, collapse is halted and reversed (while accretion rate slows-down rapidly) and the mass shells re-expand for a few $t_{\rm rh}$.

Collapse times for tidal limited models with same initial conditions, can be seen in Table~\ref{collttid}, where collapse parameters of rotating BH and non-BH models are compared. Collapse times are comparable to the non-BH models and slightly smaller for increasing rotation. The cluster loses, at collapse time, between 40 and 85\% of its mass in the non-BH models and between 48 an 87 \% of the initial cluster mass in the BH models. In both cases the mass-loss is higher for higher rotation. The time at which the cluster loses half of its mass is shorter, the higher the initial rotation of the model. The effect is driven by the interplay between relaxation effects (gravo-gyro instabilities), BH star accretion and tidal mass-loss. This will also involve a faster dissolution of the cluster in the galactic tidal field.

\begin{table}\centering
\begin{tabular}{ | c || c | c | c || c | c | c |}
\hline & \multicolumn{3}{c ||}{non-BH models}& \multicolumn{3}{c |}{BH models}\\
\cline{2-7}Model& $\frac{t_{\rm cc}}{t_{\rm rhi}}$ & $M_{\rm cc}$ & $\frac{t_{50}}{t_{\rm rhi}}$&$\frac{t_{\rm cc}}{t_{\rm rhi}}$&$M_{\rm cc}$ &$\frac{t_{50}}{t_{\rm rhi}}$ \\
\hline \hline 
M1T &  11.80 & 0.60 & 13.20 & 11.58 & 0.52 & 11.88 \\ 
M2T &  10.46 & 0.48 & 10.10 & 10.08 & 0.43 & 9.29 \\ 
M3T &  7.24 & 0.33 &  5.40 &  6.95 & 0.29 & 4.93 \\ 
M4T &  4.85 & 0.23 &  2.60 &  4.44 & 0.19 & 2.23 \\ 
M5T &  3.45 & 0.18 &  2.20 &  2.72 & 0.13 & 0.98 \\
\hline
\end{tabular}\\
\caption[Collapse parameters]{Comparison of collapse parameters between rotating BH and non-BH models. $t_{\rm cc}$: core collapse time\\ $t_{50}$: time at which the cluster lost half of its mass \\ $M_{\rm cc}$: current cluster mass at $t \approx t_{\rm cc}$}
\label{collttid}
\end{table}   

\begin{table}
  \centering
  \begin{tabular}{ | c | c | c | c | c |} \hline
      Model& $M_{\rm bh}^{\rm stall}/M{\rm i}$ & $M_{\rm bh}^{\rm stall} \ (M_\odot)$ & $\frac{d(M/M{\rm i})}{d(t/t_{\rm rhi})}_{\rm max}$ & $\frac{dM}{dt}_{\rm max} \ (\frac{M_\odot}{\rm yr})$ \\
        \hline \hline
    M1T & $1.5 \cdot 10^{-3}$ & $7.6 \cdot 10^3$ & $7.0 \cdot 10^{-4}$ & $3.5 \cdot 10^{-6}$\\ 
    M2T & $1.3 \cdot 10^{-3}$ & $6.4 \cdot 10^3$ & $8.0 \cdot 10^{-4}$ & $4.0 \cdot 10^{-6}$\\ 
    M3T & $9.3 \cdot 10^{-4}$ & $4.7 \cdot 10^3$ & $1.1 \cdot 10^{-3}$ & $5.5 \cdot 10^{-6}$\\ 
    M4T & $6.7 \cdot 10^{-4}$ & $3.3 \cdot 10^3$ & $1.5 \cdot 10^{-3}$ & $7.5 \cdot 10^{-6}$\\ 
    M5T & $4.6 \cdot 10^{-4}$ & $2.3 \cdot 10^3$ & $1.8 \cdot 10^{-3}$ & $9.0 \cdot 10^{-6}$\\ \hline
  \end{tabular}
   \caption[Description of mass parameters]{\small Description of mass parameters. Column 1: Model name; Column 2: Final $M_{\rm bh}$ in units of initial total mass as it reaches an asymptotic mass; Column 3:  $M_{\rm bh}$ in $M_\odot$-units; Column 4: maximal accretion rate $dM/dt$ in units of initial mass and half-mass relaxation time; Column 5: $dM/dt$ in $(M_\odot$/yr)-units.}
  \label{masstid}
 \end{table}

Table~\ref{masstid} shows the final BH mass for each model and the respective maximal accretion rates. $M_{\rm bh}^{\rm stall}$ varies between $2.3 \cdot 10^3 M_\odot$ and  $7.7 \cdot 10^3 M_\odot$, which are smaller masses than in the isolated case. The general behavior exhibits a decreasing $M_{\rm bh}^{\rm stall}$ but a higher mass growth rates corresponding to higher initial rotation. A stalling of $M_{\rm bh}$ is always faster the higher $\omega_0$ is.

The cluster mass-loss due to tidal effects of the parent galaxy is very strong during the re-expansion of the core. The acceleration of mass-loss is similar as the observed by \cite{kim02} in the post-collapse models driven by binary heating, although the effect in the present BH-models is more pronounced, with the consequence of a faster evolution of the cluster towards sphericity and final dissolution.

Tidally limited models experience the onset of gravo-gyro instabilities, as angular momentum diffuses outwards, leading to a strong mass-loss and a experience a limited increase of central rotation. At the galaxy tidal boundary, mainly circular tidal orbits in the halo are lost, while high eccentric (low $J_z$) orbits can interact with stars in the core.

\begin{figure}
\includegraphics[width=\linewidth]{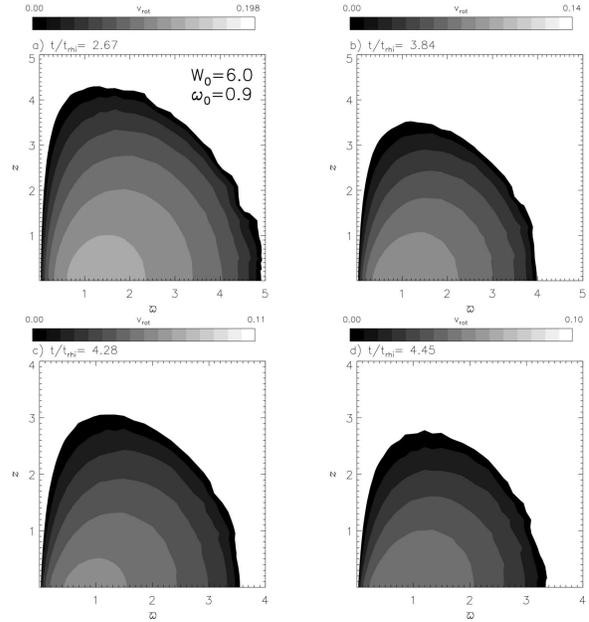}
\caption{\small 2D rotational velocity distribution in the meridional plane for a model M4T ($6.0,0.9$) in the tidally limited case.}
\label{vrotmtime060090tid}
\end{figure}

Fig.~\ref{vrotmtime060090tid} shows snapshots of the evolution of 2-dimensional distribution of rotational velocity in the meridional plane, at representative times, where the lighter areas represent contours of higher rotation. Rotation is lost stronger than in the isolated model. Angular momentum transport and the growing BH-potential support the development of ordered motion in the core and, at the same time, trigger mass-loss through the tidal boundary.

\section{Conclusions and outlook}
\label{sec:6}
The variety of environments in which dense stellar systems (GCs, GN) form and evolve, make them target of study of fundamental dynamical processes. Observational studies of GCs suggest the existence of IMBHs in their centers (like G1 and M15) and is well known, that some of them show flattening due to system rotation.
The improvements in observational and theoretical studies of dense stellar systems in the last years, has led to a better understanding of central BHs and their environments, but at the same time, opened new questions due to their complexity. This is the reason, why theoretical models are of importance to elucidate the origin of the observed phenomena, to be able to explain their formation, evolution and interaction, and predict possible evolutionary scenarios, which can be confirmed by observations. The presented theoretically formulated evolutionary models extend the model complexity of spherically symmetric systems through the implementation of differential rotation and BH star accretion.

Results can be summarized as follows:
\begin{itemize}
\item  Core collapse is an evolutionary property in self-gravitating systems without BH \citep{einsel99,kim02}. Gravo-gyro effects are coupled to gravothermal instability and drive core contraction. We start with a seed central BH, which grows over relaxation time scales due to stellar accretion. Evolution towards core contraction increases the central density and supports star accretion by the central BH. The BH acts as an energy source, through which energetic stellar orbits are formed, which easily go into the loss-cone or interact with stars inside $r_{\rm a}$ being able to reverse collapse, triggering core expansion. A rapid expansion has been observed as well in $N$-Body realizations \citep{baum04}, which use high concentrated initial models ($W_0=10$) and higher initial BH masses ($\sim 1-5 \% M_{\rm cl}$) in their simulations. They show accretion rates which agree to the classical approximation \citep{fr76} applied to a Bahcall-Wolf cusp, as we also show in Figs.~\ref{timemassisocomp} and \ref{timemassisocompalpha}. In the presence of an external potential (tidal limit), a faster evolution accelerates mass-loss and leads to cluster dissolution.

\item Final steady state solutions are found in all isolated models, which approach the -1.75 slope in the density cusp and the -0.5 slope in the velocity dispersion cusp inside the BH influence radius $r_{\rm a}$, corresponding to Keplerian bounded orbits, independent of initial rotation.
\item The final $M_{\rm bh}$ nearly stalls at $\sim 0.001 M_{\rm cli}$, and grows slower during the post-collapse phase, while BH mass accretion rate ($dM_{\rm bh}/dt$) decreases strongly after reaching a maximum before core expansion, due to the higher density of orbits in the core. In the tidally limited model mass-loss is very strong during the re-expansion of the core. The cluster mass reaches in a fraction of $t_{\rm rh_i}$ after collapse values at least one order of magnitude smaller than its hosted BH. For a cluster of $5 \cdot 10^6 M_\odot$,  $M_{\rm bh}^{\rm stall}$ varies between $7.5 \cdot 10^3 M_\odot$ and  $1.5 \cdot 10^4 M_\odot$ for isolated models of different initial rotation, and between $2.3 \cdot 10^3 M_\odot$ and  $7.6 \cdot 10^3 M_\odot$ for tidally limited models.
As mentioned in Sect.~\ref{sec:31} this values reproduce the expected mass of IMBHs, since we set our initial parameters, according to physical properties of these objects.
\item High rotating, moderate concentrated models (M4, M5) maintain central rotation at collapse and during the expansion phase in an efficient way, in comparison with models without BH. They are able to maintain an efficient angular momentum diffusion, and at the same time are concentrate enough to avoid an excessive mass-loss. Both effects support the accretion of stars in low-$J_{\rm z}$ orbits. These models show a stable evolution of ordered vs. random motion ($V_{\rm rot} / \sigma$) in BH-models, up to the expansion phase.
\end{itemize}

Since flattening supported by rotation is a well known phenomena in GCs and observational evidences of the existence of central dark objects in some GCs, with and without rotation, there is a motivation for the study of this constraint in the long term evolution.
Although some constraints are still missing, like a mass spectrum or stellar evolution (work in progress), as well as a more realistic criterion for tidal mass-loss (as observations suggest, e.g. \citealt{mackey05}), our models are consistent with existent theoretical studies on the general evolution of systems embedding BHs and additionally consider the importance of initial differential rotation, which needs to be taken into account for the understanding of GC formation and evolution, specially when it can be high enough in young clusters (e.g. in the LMC, \cite{brocato04}).

Moreover, galaxy cores are known to harbor SMBHs and some of them are 'collisional', in the sense that their relaxation times are $ \lesssim 10^{10}$ yr (like M32 and the MW). They show cuspy density profiles, which shorten the times of relaxation the smaller the distance to the center of the nuclei and might support the formation of a steady state configuration (Bahcall-Wolf solution) at times of the order of 10 Gyrs.

Since the models presented here have only one mass component, the effect of mass segregation in a realistic multi-mass system will shorter times of evolution \citep{gurkan04}, leading to a faster set-off of expansion in the presence of a central BH. The central velocity dispersion arises because of the BH-induced Bahcall-Wolf cusp. This increase is not affected by the presence of a spectrum of stellar masses but the high mass stars are expected to show a lower cusp in their central dispersion (as reported by \citealt{kim04}), leading possibly to a higher or at least more stable $V_{\rm rot} / \sigma$ for this mass classes. Multi-mass models with BH are being currently developed and comparison to $N$-Body models are aimed to complement this calculations, using the highest particle number permitted at the time ($N \sim 10^6$) \citep{berczik06}.

As shown, the amount of rotation present in the system during its dynamical evolution is strong influenced by the interplay between angular momentum diffusion (gravo-gyro instability) and the redistribution of high energy orbits close to the BH (loss-cone refilling). Since a central BH is able to 'consume' angular momentum from the system, in form of stars, it might become itself an angular momentum source, which could be able to rotate (Kerr Black Hole), permitting also a more efficient angular momentum transport outwards, through interaction with core stars, driven by relaxation. A binary black hole (BBH), could in a similar way, lead to a more efficient support in the development of rotation in its zone of influence, modifying substantially the final shape of the cluster \citep{mapelli05,berczik06}.

The models presented, are able to reproduce 2D distributions (in the meridional plane) of density, cluster- and BH-potential, velocity dispersions, rotational velocity, anisotropy, dynamical ellipticity, among other parameters, at any time of evolution and deep in the stellar cusp surrounding the central BH. They make possible the study of kinematical and structural parameters in time, which can complement and test observational measurements contributing to the understanding of the common evolution of star clusters and galaxies.

\section*{Acknowledgments}
We thank the referee for fruitful comments which helped to improve the quality of the paper. We are grateful to H.M. Lee and E. Kim for helpful discussions. This work was supported through grants Sp 345/17-1 and 17-2 in the framework of
priority program SPP 1177 of German Science Foundation (DFG).

\label{lastpage}

\end{document}